\documentclass[11pt,a4paper,aps,article,nofootinbib]{revtex4-2}

\usepackage[utf8]{inputenc}
\usepackage[T1]{fontenc}
\usepackage{graphicx, caption, subcaption}  
\usepackage{dcolumn}   
\usepackage{amsmath, amssymb}       
\usepackage{color,latexsym,array,multirow,verbatim,enumerate}
\usepackage{url}
\usepackage{cancel}
\usepackage{hyperref}
\usepackage{mathrsfs}

\begin{document}
	\title{VERIFIABLE  TYPE-II  SEESAW  AND  DARK MATTER IN \\ A GAUGED $U(1)_{B-L}$ MODEL}
	\author{Satyabrata Mahapatra\footnote{Email:PH18RESCH11001@iith.ac.in}}
	\author{Nimmala Narendra\footnote{Email:PH14RESCH01002@iith.ac.in}} 
	\author{Narendra Sahu\footnote{Email:nsahu@iith.ac.in}}
	\affiliation{Indian Institute of Technology Hyderabad, Kandi, Sangareddy, 502285, Telangana, India.}
	
	\begin{abstract} We propose a gauged $U(1)_{B-L}$ extension of the standard model (SM) to explain simultaneously 
		the light neutrino masses and dark matter (DM). The generation of neutrino masses occurs through a variant 
		of type-II seesaw mechanism in which one of the scalar triplets lies at the TeV scale yet have a large 
		dilepton coupling, which paves a path for probing this model at colliders. The gauging of $U(1)_{B-L}$ symmetry 
		in a type-II seesaw framework introduces $B-L$ anomalies. Therefore we invoke three right handed neutrinos 
		$\nu_{R_{i}}$(i=1,2,3) with $B-L$ charges -4,-4,+5 to cancel the anomalies. We further show that the lightest 
		one among the three right handed neutrinos can be a viable DM candidate. The stability of DM can be owed to a 
		remnant $Z_2$ symmetry under which the right handed neutrinos are odd while all other particles are even. We 
		then discuss the constraints on the model parameters from observed DM abundance and the search at direct 
		detection experiments.  
	\end{abstract}	
		\maketitle{}
		\section{Introduction}\label{intro}
		The Standard Model (SM) of Particle physics works remarkably well in describing the electroweak and strong interactions 
		of fundamental particles of nature. But there are many questions which are unanswered till date. Among them the most 
		popular unsolved problems are the identity of Dark Matter (DM) and the origin of neutrino masses. The cosmological 
		observations like the galaxy rotation curve, gravitational lensing and large scale structure of the Universe provide 
		the evidences towards the existence of DM~\cite{Bertone:2004pz,Feng:2010gw}. But we do not have much information about 
		microscopic properties of DM apart from its relic density, which is precisely measured by the WMAP~\cite{Hinshaw:2012aka} and 
		PLANCK~\cite{Aghanim:2018eyx} to be $\Omega_{\rm DM}h^{2}=0.120\pm 0.001$.
		
		Initially the neutrinos were thought to be massless particles due to lack of experimental evidences. But the neutrino 
		oscilation experiments~\cite{solar-expt,atmos-expt,kamland} confirmed that they are massive but tiny. Assuming that the 
		neutrinos are Majorana ({$\triangle L=2$}), their sub-eV masses are best understood by the dimension five operator: 
		$\mathcal{O}_{5}=\frac{LLHH}{\Lambda}$ , where $L$ and $H$ are the lepton and Higgs doublets of the SM and $\Lambda$ is 
		the scale of new physics~\cite{Weinberg:1979sa,Ma:1998dn}. After electroweak phase transition, we get sub-eV neutrino 
		masses $M_{\nu}  =\mathrm{O}(\frac{{\langle H \rangle}^2}{{\Lambda}}) \simeq 0.1$ eV, for $<H>=10^{2}$GeV and $\Lambda=10^{14}$GeV. 
		Seesaw mechanisms: Type-I~\cite{Minkowski:1977sc,GellMann:1980vs,Mohapatra:1979ia,Schechter:1980gr}, Type-II~\cite{Mohapatra:1980yp,Lazarides:1980nt,Schechter:1981cv,Wetterich:1981bx,Brahmachari:1997cq} and Type-III~\cite{Foot:1988aq} are 
		the various UV completed realizations of this dimension five operator. In the type-I seesaw one introduces heavy singlet 
		RHNs while in case of type-II and type-III, one introduces a triplet scalar($\Delta$) of hypercharge 2 and triplet fermions 
		of hypercharge 0 respectively. 
		
		In the conventional type-II seesaw, the relevant terms in the Lagrangian which violates lepton number by two units are  
		$f_{\alpha\beta}\Delta L_{\alpha} L_{\beta} + \mu \Delta^{\dagger}HH$, where $\Delta$ does not acquire an explicit vacuum 
		expectation value(vev). However, the electro-weak phase transition induces a small vev of $\Delta$ as: $\langle \Delta \rangle 
		= -\frac{\mu  \langle H \rangle^2}{M^2_{\Delta}}$. Thus for $\mu \sim M_{\Delta} \sim 10^{14}$GeV, one can get $M_{\nu}=f \langle \Delta \rangle 
		\simeq f\frac{\langle H \rangle^2}{M_{\Delta}}$ of order  $\mathrm{O}$(0.1)eV for $f\sim 1$. The only drawback in this case is that the mass 
		scale of the scalar triplet is much larger than the energy attainable at current generation colliders. Hence such models 
		lack falsifiability.  
		
		Alternatively one can introduce two scalar triplets: $\Delta$  and  $\xi$ with $M_{\Delta} \sim 10^{14}$GeV and $M_{\xi} \sim$ TeV 
		$<<$ $M_{\Delta}$\cite{McDonald:2007ka} \footnote{A modified double type-II seesaw with TeV scale scalar triplet is also proposed in 
			ref. \cite{Gu:2009hu}}. If one imposes an additional $B-L$ gauge symmetry~\cite{Majee:2010ar}, then the relevant terms 
		in the Lagrangian are: $\mu \Delta^{\dagger}HH + f\xi LL + y\Phi^2_{ B-L}\Delta^{\dagger} \xi$, where $\Phi_{ B-L}$ is the scalar 
		field responsible for $B-L$ symmetry breaking. At TeV scales $\Phi_{ B-L}$ acquires a vev and break $B-L$ symmetry. Moreover 
		$\langle \Phi_{ B-L} \rangle$ generates a small mixing between $\Delta $ and $\xi$ of the order $\theta \sim \frac{\langle \Phi_{ B-L}\rangle^{2}}
		{M^2_{\Delta}} \simeq 10^{-18}$. This implies that $\xi LL$ coupling can be large while $\xi$'s coupling with Higgs doublet is 
		highly suppressed. Since $\Delta$ mass is super heavy, it gets decoupled from the low energy effective theory. On the other hand, 
		$\xi$ can be at TeV scale with large dilepton coupling. As a result the same sign dilepton signature of $\xi$ can be studied at 
		colliders~\cite{Chun:2003ej,Perez:2008ha,Chun:2019hce,Padhan:2019jlc,Akeroyd:2005gt,Huitu:1996su,Hektor:2007uu,Mitra:2016wpr}.  
		
		The gauging of $U(1)_{\rm B-L}$ symmetry in a type-II seesaw framework introduces non-trivial gauge and gravitational anomalies. With 
		the SM particle content all triangle anomalies are zero except for ${\sum} {[U(1)_{{B-L}}]}^3=3$ and  $\sum [Grav.]^2 \times [U(1)_{{B-L}}]=3$. 
		These anomalies can be cancelled by introducing new fermions in such a way that sum of their ${{B-L}}$ quantum numbers is $-3$. In this 
		paper we introduce three right handed neutrinos $\nu_{R_{i}}$ $(i=1,2,3)$ with $U(1)_{{B-L}}$ charges $-4, -4, +5$ respectively, such that 
		$\sum_{i=1}^{3} (Y_{{B-L}})_{i}=-3$, to make the theory anomaly free~\cite{Montero:2007cd,Sanchez-Vega:2014rka,Ma:2014qra,Sanchez-Vega:2015qva,
			Ma:2015mjd,Patra:2016ofq,Nanda:2017bmi,Gu:2019ird}. Interestingly, one of the three right handed neutrions can be a viable candidate of DM. 
		The stability of the DM candidate can be guaranteed by a remanant $Z_{2}$ symmetry of the original $U(1)_{B-L}$. Under the $Z_{2}$ 
		discrete symmetry $\nu_{R_{i}}\,(i=1, 2, 3)$ are odd while all other particles are even. Thus without imposing any additional discrete 
		symmetry we can explain the DM as well as smallness of the neutrino masses.
		
		The paper is organised as follows. In section 2, we briefly discuss the gauge and gravitaional anomalies and hence the anomaly free 
		conditions of a gauged $U(1)_{\rm B-L}$ extension of a type-II seesaw model. In section 3, we describe the proposed model, the scalar 
		masses and mixing and the neutrino mass generation at TeV scale through a variant of type-II seesaw. We then discuss how the 
		particles introduced for anomaly cancellation become viable DM candidate and study the relic density and it's compatibility 
		with the direct detection experiments in section 4. We also briefly discuss the collider signatures of the model in section 5 and 
		finally conclude in section 6.

		\section{Anomalies in a gauged $U(1)_{\rm B-L}$ extension of a type-II seesaw}\label{anomaly}
		Within the SM, $U(1)_{\rm B-L}$ is happend to be an accidental global symmetry. However, the gauged $U(1)_{\rm B-L}$ extension of the SM is 
		not anomaly free. Among all the anomalies arising from the triangle diagrams involving the gauge currents, except $\mathcal{A}[U(1)^3_{B-L}]$ and 
		$\mathcal{A}[(Gravity)^2 \times U(1)_{B-L} ]$, all others are trivial. Here $\mathcal{A}$ stands for the anomaly coefficient which 
		in a chiral gauge theory is given by~\cite{pbpalbook}:
		\begin{equation}
		\mathcal{A}=Tr[T_a[T_b,T_c]_+]_R-Tr[Ta[T_b,T_c]_+]_L\,,
		\end{equation} 
		where the $T$ denotes the generators of the gauge groups and $R$ and $L$ represents the interactions of right and left chiral fermions with the gauge bosons.
		
		Gauging of $U(1)_{\rm B-L}$ symmetry within the SM lead to the following triangle anomalies: 
		\begin{equation*}
		\mathcal{A}_1[U(1)^3_{B-L}]=3
		\end{equation*}
		\begin{equation}
		\mathcal{A}_{2}[(Gravity)^2 \times U(1)_{B-L}]=3\,.
		\end{equation}
		If three right handed neutrinos, each of having $B-L$ charge $-1$, are added to the SM, then they result in $\mathcal{A}_1[U(1)^3_{B-L}]=-3$ 
		and $\mathcal{A}_{2}[(Gravity)^2 \times U(1)_{B-L}]=-3$ which lead to cancellation of above mentioned gauge anomalies. This is the most natural 
		choice to make $U(1)_{B-L}$ models anomaly free. However we can have alternative ways of costructing anomaly free versions of $U(1)_{B-L}$ extension 
		of the SM~\cite{Montero:2007cd,Sanchez-Vega:2014rka,Ma:2014qra,Sanchez-Vega:2015qva,Ma:2015mjd,Patra:2016ofq,Nanda:2017bmi,Gu:2019ird}. In particular, 
		three right handed neutrinos with exotic $B-L$ charges -4,-4,+5 can also give rise to vanishing $B-L$ anomalies as follows.
		\begin{equation*}
		\mathcal{A}_1[U(1)^3_{B-L}]=\mathcal{A}^{SM}_1[U(1)^3_{B-L}]+\mathcal{A}^{New}_1[U(1)^3_{B-L}]= 3+[(-4)^3+(-4)^3+(5)^3]=0
		\end{equation*}
		\begin{align}
		\mathcal{A}_{2}[(Gravity)^2 \times U(1)_{B-L}]&=\mathcal{A}^{SM}_{2}[(Gravity)^2 \times U(1)_{B-L}]
		+\mathcal{A}^{New}_{2}[(Gravity)^2 \times U(1)_{B-L}]\nonumber\\ &=3+[(-4)+(-4)+(5)]=0
		\label{anomaly_cancel}
		\end{align} 
		
		In a type-II seesaw framework, the SM is usually extended with a triplet scalar of hypercharge 2. In this case, gauging of $U(1)_{\rm B-L}$ symmetry 
		does not lead to any new anomalies apart from the mentioned above. Therefore, in what follows, we consider a type-II seesaw framework 
		with gauged $U(1)_{\rm B-L}$ symmetry, where the $B-L$ anomalies are cancelled by the introduction of three right handed neutrinos with 
		exotic $B-L$ charges -4,-4,+5. The model thus proposed explains the origin of neutrino mass and DM in a minimal set-up.

		\section{The Complete Model}\label{model}
		We study a variant of type-II seesaw model based on the gauge group $SU(3)_C \times SU(2)_L \times U(1)_Y \times U(1)_{{B-L}}$, where ${B}$ and ${L}$ 
		stands for the usual baryon and lepton numbers, respectively. Two triplet (under $SU(2)_{L}$) scalars $\Delta$ and $\xi$: 
		\begin{equation}
		\Delta = \begin{pmatrix} \frac{\delta^+}{\sqrt{2}} &&  \delta^{++}\\
		\delta^0  && -\frac{\delta^+}{\sqrt{2}} \\
		\end{pmatrix}  
		~~~{\rm and}~~~
		\xi =\begin{pmatrix}
		\frac{\xi^{+}}{\sqrt{2}} && \xi^{++}\\
		\xi^0 &&  -\frac{\xi^+}{\sqrt{2}} \\
		\end{pmatrix}
		\end{equation}  
		with $M_{\Delta} \sim 10^{14}$ GeV and $M_{\xi} \sim$ TeV $<<$ $M_{\Delta}$ are introduced, where the $B-L$ charges of $\Delta$ and $\xi$ are 0 and 2 
		respectively. As discussed in the previous section, the additional $U(1)_{B-L}$ gauge symmetry introduces $B-L$ anomalies in the theory. To cancel these 
		$B-L$ anomalies we introduce three right handed neutrinos $\nu_{R_{i}}$ $(i=1, 2, 3)$, where the $B-L$ charges of $\nu_{R_{1}}$, $\nu_{R_{2}}$, 
		$\nu_{R_{3}}$ are -4, -4, +5 respectively as shown in Eq.\ref{anomaly_cancel}. Note that such unconventional $B-L$ charge assignment of the $\nu_{R_{i}}\,(i=1,2,3)$ forbids their Yukawa couplings with the SM particles. We also introduce three singlet scalars: $\Phi_{B-L}$, $\Phi_{12}$ and $\Phi_{3}$ whose ${B-L}$ charges 
		are given by: -1, +8, -10 respectively. As a result $\Phi_{12}$ and $\Phi_{3}$ couples to $\nu_{R_{1,2}}$ and $\nu_{R_{3}}$ respectively through Yukawa terms. 
		The vev's of $\Phi_{12}$ and $\Phi_{3}$ provides Majorana masses to the right handed neutrinos, while the vev of $\Phi_{B-L}$ provides a small mixing between 
		$\Delta$ and $\xi$. We will show later that the mixing between $\Delta$ and $\xi$ plays a key role in generating sub-eV masses of neutrinos. The particle 
		content and their charge assignments are listed in Table.\ref{table}.
		
		\begin{table}
			\begin{center}
				\begin{tabular}{|c c|c|c|c|c|c|}
					\hline
					& Fields & $SU(3)_c$ & $SU(2)_L$ & $U(1)_Y$ & $U(1)_{{B-L}}$\\ 
					\hline
					& $\nu_{R_{1,2}}$ & 1 & 1 & 0 & -4 \\
					& $\nu_{R_{3}}$ & 1 & 1 & 0 & 5 \\
					\hline
					& $\Delta$ & 1 & 3 & 2 & 0 \\
					& $\xi$ & 1 & 3 & 2 & 2 \\
					& $\Phi_{{B-L}}$ & 1 & 1 & 0 & -1 \\
					& $\Phi_{12}$ & 1 & 1 & 0 & +8 \\
					& $\Phi_3$ & 1 & 1 & 0 & -10 \\
					\hline
				\end{tabular}   
				\caption{New particles and their quantum numbers under the imposed gauge symmetry.}
				\label{table}
			\end{center}
		\end{table}
		
		The Lagrangian involving the new fields consistent with the extended symmetry can be written as:
		\begin{align}
		\mathscr{L} &\supset \overline{\nu_{R_{a}}}\, i \gamma^\mu D_{\mu} \nu_{R_{b}} + \overline{\nu_{R_{3}}} i \gamma^\mu D_{\mu} \nu_{R_{3}} + {\big| D_{\mu} X \big|}^2 \nonumber\\
		&   + Y_{ij}^{\xi}~ \overline{L_{i}^c}\,\, i \tau_{2} \, \xi \, L_{j} + Y_{ab}\,\, \Phi_{12}\,\, \overline{(\nu_{R_{a}})^c}\,\, \nu_{R_{b}} +Y_{33}\,\, \Phi_{3}\,\, \overline{(\nu_{R_{3}})^c}~ \nu_{_{R3}}  \nonumber\\
		&   + Y_{a3} \overline{(\nu_{R_{a}})^c}~ \Phi_{_{B-L}} \nu_{R_{3}}  + {\rm h.c.} 
		- \mathrm{V}(H,\Delta,\xi,\Phi_{{B-L}}, \Phi_{12}, \Phi_3)
		\label{full_Lagr}
		\end{align}
		where 
		\begin{equation}
		D_{\mu}=\left( \partial_\mu + i~g_{_{_{B-L}}} Y_{_{B-L}}  (Z_{_{B-L}})_\mu \right) \nonumber\\.
		\end{equation}
		The $g_{_{B-L}}$ is the gauge coupling associated with $U(1)_{B-L}$ and $Z_{B-L}$ is the corresponding gauge boson. Here, $X=  \Phi_{{B-L}}, \Phi_{12}, \Phi_3$. The $i,j$ runs over 1, 2, 3 and $a,b$ runs over 1,2.
		
		
		The scalar potential of the model can be written as:
		\begin{align}\label{scalar_potential}
		\mathrm{V}(H,\Delta,\xi,\Phi_{{B-L}}, \Phi_{12}, \Phi_3) &= -{\mu}^2_H H^\dagger H + \lambda_H {(H^\dagger H)^2}+ M^2_{\Delta} \Delta^\dagger \Delta + \lambda_{\Delta} {(\Delta^\dagger \Delta)^2} \nonumber\\
		&  + M^2_{\xi} \xi^\dagger \xi + \lambda_{\xi} {(\xi^\dagger \xi)^2} -  {\mu}^2_{\Phi_{{B-L}}} \Phi_{{B-L}}^\dagger  \Phi_{{B-L}} + \lambda_{\Phi_{{B-L}}} (\Phi_{{B-L}}^\dagger  \Phi_{{B-L}})^2 \nonumber\\
		&  - {\mu}^2_{\Phi_{12}} \Phi_{12}^\dagger \Phi_{12} + \lambda_{\Phi_{12}} {(\Phi_{12}^\dagger \Phi_{12})^2} - {\mu}^2_{\Phi_3} \Phi^\dagger_3 \Phi_3 + \lambda_{\Phi_3} {(\Phi^\dagger_3 \Phi_3)^2} \nonumber\\
		&  + \lambda_{H \Delta} (H^\dagger H)(\Delta^\dagger \Delta) + \lambda_{H \xi} (H^\dagger H)(\xi^\dagger \xi) \nonumber\\
		&  +\lambda_{H \Phi_{{B-L}}} (H^\dagger H)(\Phi_{{B-L}}^\dagger  \Phi_{{B-L}}) + \lambda_{H \Phi_{12}} (H^\dagger H)({\Phi_{12}}^\dagger \Phi_{12}) \nonumber\\
		&  + \lambda_{H \Phi_3} (H^\dagger H)(\Phi^\dagger_3 \Phi_3) + \lambda_{\Delta \xi} (\Delta^\dagger \Delta) (\xi^\dagger \xi) + \lambda'_{\Delta \xi} (\Delta^\dagger \xi) (\xi^\dagger \Delta) \nonumber\\
		&  + \lambda_{\Delta \Phi_{{B-L}}} (\Delta^\dagger \Delta) (\Phi^\dagger_{{B-L}} \Phi_{{B-L}}) + \lambda_{\Delta \Phi_{12}} (\Delta^\dagger \Delta) ({\Phi_{12}}^\dagger \Phi_{12}) \nonumber\\
		&  + \lambda_{\Delta \Phi_3} (\Delta^\dagger \Delta) (\Phi^\dagger_3 \Phi_3) + \lambda_{\xi \Phi_{{B-L}}} (\xi^\dagger \xi)(\Phi^\dagger_{{B-L}} \Phi_{{B-L}})\nonumber\\
		&  + \lambda_{\xi \Phi_{12}}(\xi^\dagger \xi)({\Phi_{12}}^\dagger \Phi_{12}) + \lambda_{\xi \Phi_3}(\xi^\dagger \xi)(\Phi^\dagger_3 \Phi_3) \nonumber\\
		&  +\lambda_{\Phi_{{B-L}}\Phi_{12}} (\Phi_{{B-L}}^\dagger  \Phi_{{B-L}})({\Phi_{12}}^\dagger \Phi_{12}) + \lambda'_{\Phi_{{B-L}}\Phi_{12}} (\Phi_{{B-L}}^\dagger  \Phi_{12})({\Phi_{12}}^\dagger \Phi_{B-L})\nonumber\\
		&  + \lambda_{\Phi_{{B-L}}\Phi_3} (\Phi_{{B-L}}^\dagger  \Phi_{{B-L}})(\Phi^\dagger_3 \Phi_3)+ \lambda'_{\Phi_{{B-L}}\Phi_3} (\Phi_{{B-L}}^\dagger  \Phi_{3})(\Phi^\dagger_3 \Phi_{B-L})\nonumber\\
		&  + \lambda_{\Phi_{12} \Phi_3}({\Phi_{12}}^\dagger \Phi_{12})({\Phi}^\dagger_3 \Phi_3)+ \lambda'_{\Phi_{12} \Phi_3}({\Phi_{12}}^\dagger \Phi_3)(\Phi^\dagger_3 \Phi_{12})+\mu \Delta^\dagger H H\nonumber\\
		&  + y \Phi_{{B-L}}^2 \Delta^\dagger \xi + Y (\Phi_{{B-L}}^\dagger)^2 \Phi_3 \Phi_{12} + h.c.
		\end{align}
		
		It is worth mentioning that the mass squared terms of $\Delta$ and $\xi$ are chosen to be positive so that only the neutral components of 
		$H$,$\Phi_{12}$,$\Phi_{ B-L}$ and $\Phi_3$ acquire non-zero vevs. These vevs are given as follows:
		\begin{equation*}
		\langle H \rangle=\frac{v}{\sqrt{2}}\begin{pmatrix}
		0\\ 1
		\end{pmatrix},  \langle \Phi_{12} \rangle =\frac{v_{_{12}}}{\sqrt{2}},  \langle \Phi_{ B-L} \rangle =\frac{v_{_{B-L}}}{\sqrt{2}} ,   
		\langle \Phi_{3} \rangle=\frac{v_{_3}}{\sqrt{2}}\,.
		\end{equation*}
		
		However, after electroweak phase transition, $\Delta$ and $\xi$ acquire induced vevs:
		\begin{equation*}
		\langle \Delta \rangle =\frac{u_{_\Delta}}{\sqrt{2}}\begin{pmatrix}
		0 && 0\\
		1 && 0\\
		\end{pmatrix}  {\rm and}~~
		\langle \xi \rangle=\frac{u_{_\xi}}{\sqrt{2}}\begin{pmatrix}
		0 && 0\\
		1 &&  0\\
		\end{pmatrix}
		\end{equation*}  
		
		\subsection{Neutrino Mass}\label{neutrinomass}
		
\begin{figure}[h!]
					\begin{center}
						\includegraphics[scale=0.3]{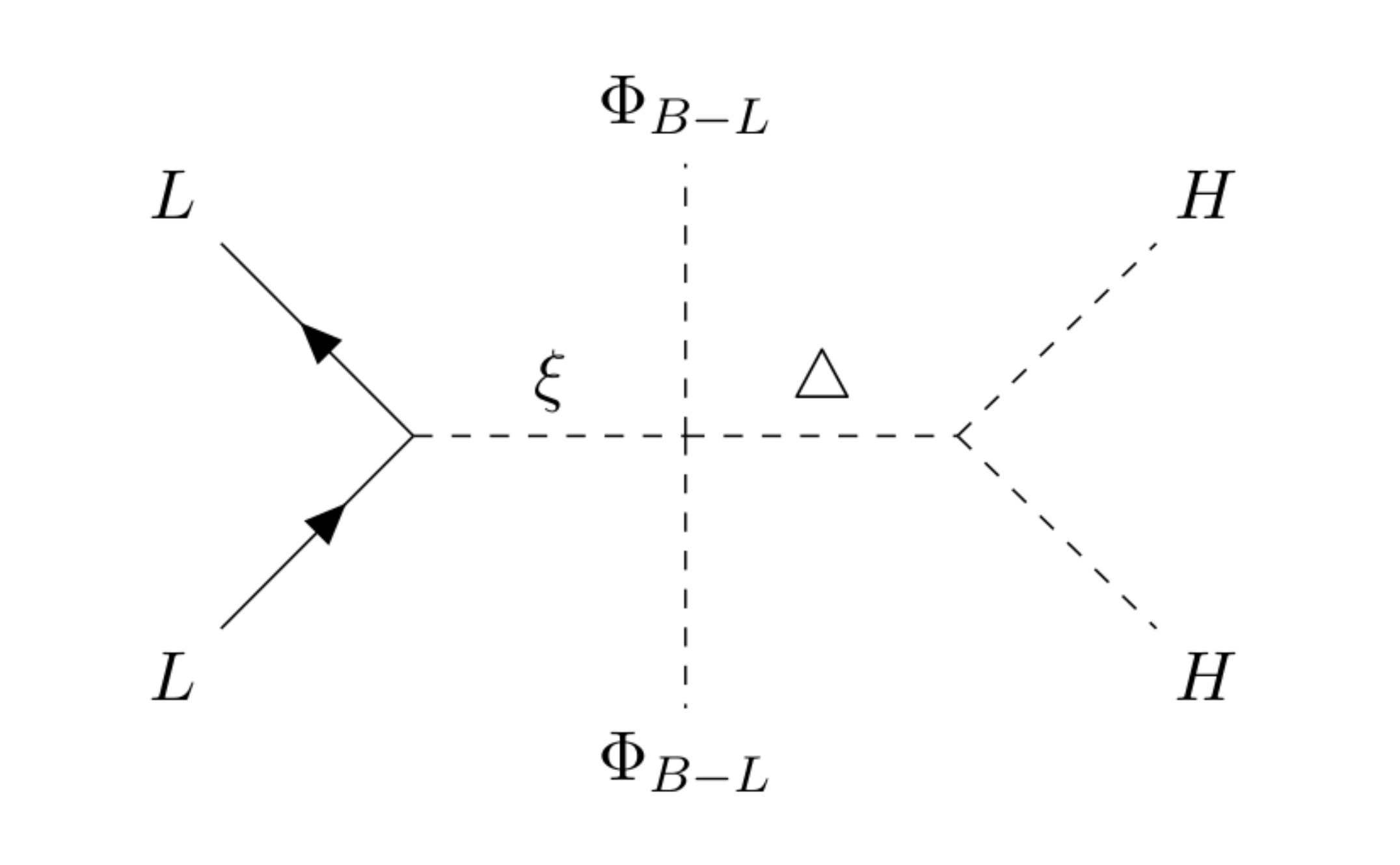}
				
				\caption{\footnotesize{Generation of neutrino mass through the modified Type-II seesaw.}}
				\label{neutrino_mass}
			\end{center}
		\end{figure}

		Until $\Phi_{{B-L}}$ gains a vev, there is no mixing between $\Delta$ and $\xi $. Once $\Phi_{{B-L}}$ acquires a vev, say $v_{_{B-L}}$ at TeV scale, it 
		allows a mixing between $\Delta$ and $\xi $ through $y v^2_{_{B-L}} \Delta^\dagger \, \xi$. After electroweak symmetry breaking $\Delta$ gets an induced 
		vev, similarly as in the case of traditional type-II seesaw. Since $B-L$ quantum number of $\Delta$ is zero, it does not generate Majorana masses of neutrinos. 
		The induced vev acquired by $\Delta$ after EW phase transition is given by
		\begin{equation}
		\langle \Delta \rangle = u_{_\Delta} = - \frac{\mu v^2}{{\sqrt{2}}(2 M^2_\Delta + \lambda_{H \Delta}v^2 + \lambda_{\Phi_{{B-L}}\Delta} v^2_{_{B-L}}+ \lambda_{\Phi_{12} \Delta }v^2_{12}+ \lambda_{\Phi_3 \Delta }v^2_{3})}\,.
		\end{equation}
		The vev of $\Delta$ is required to satisfy: 
		\begin{equation}
		\rho \equiv \frac{M_W^2}{M_z^2 \cos^2 \theta}=\frac{1 + 2 x^2}{1 + 4 x^2} \approx 1
		\end{equation}
		where $x= u_\Delta/v$. The above constraint implies that $|u_\Delta| < \mathcal{O}(1)$ GeV. Since $\xi$ mixes with $\Delta$ after $U(1)_{B-L}$ breaking, 
		it also acquire an induced vev after EW symmetry breaking and is given by
		\begin{equation}
		\langle \xi \rangle = u_{_\xi} =  -\frac{y v^2_{_{B-L}}}{2(2 M^2_\xi +\lambda_{\xi \Phi_{{B-L}}}v^2_{_{B-L}} + \lambda_{\xi H }v^2+\lambda_{\xi \Phi_{12} }v^2_{12} + \lambda_{\xi \Phi_3 }v^2_{3})} u_\Delta\,.
		\end{equation}
		If we assume $y v^2_{{B-L}} = M^2_\xi \sim \lambda_{\xi \Phi_{{B-L}}}v^2_{_{B-L}} \sim \lambda_{\xi H }v^2 \sim \lambda_{\xi \Phi_{12} }v^2_{12} \sim \lambda_{\xi \Phi_3 }v^2_{3}$, then we get $u_\Delta \simeq u_\xi$, even if there is orders of magnitude difference in the masses of $\xi$ and $\Delta$. 
		Here we additionally assume that $M_\Delta >> v_{\rm B-L}, v_{12} >  M_\xi, v_3, v$. 
		After integrating out the heavy degrees of freedom in the Feynman diagram given in Fig.\ref{neutrino_mass}, we get the Majorana mass matrix of the light neutrinos to be
		\begin{equation}
		(M_\nu)_{ij} = Y_{ij}^{\xi} u_\xi = Y_{ij}^{\xi}~ \frac{y v^2_{{B-L}}}{2(2 M^2_\xi +\lambda_{\xi \Phi_{{B-L}}}v^2_{_{B-L}} + \lambda_{\xi H }v^2 + \lambda_{\xi \Phi }v^2_{12} + \lambda_{\xi \Phi_3 }v^2_{3})} u_\Delta .
		\end{equation}
		
		
%
%
		As $u_\Delta \simeq u_\xi \lesssim \mathcal{O}(1)$ GeV, we get sub-eV neutrino masses as required by the oscillation experiments. Here it is worth mentioning 
		that the mixing between the super heavy triplet scalar $\Delta$ and the TeV scale scalar triplet $\xi$ gives rise to the neutrino mass and it is clear from 
		Fig. \ref{neutrino_mass} that $\xi$ strongly couples to leptons while it's coupling with SM Higgs doublet is highly suppressed because of large $M_{\Delta}$.

		\subsection{Scalar Masses and Mixing}\label{scalars}
		We parametrize the neutral scalars as:
		\begin{equation*}
		H^{0}= \frac{v+h+ i\, S}{\sqrt{2}}
		~~~{\rm,}~~~
		\Phi_{B-L}=\frac{v_{_{B-L}}+\phi_{_{B-L}}+ i P}{\sqrt{2}}
		\end{equation*}
		\begin{equation*}
		\Phi_{12}=\frac{v_{12}+\phi_{12}+i\,Q}{\sqrt{2}}
		~~~{\rm,}~~~
		\Phi_{3}=\frac{v_3+\phi_3 + i\, R}{\sqrt{2}}
		\end{equation*}
		\begin{equation}
		\delta^0=\frac{u_{\Delta}+\delta + i \eta}{\sqrt{2}}		
		~~~{\rm,}~~~
		\xi^0=\frac{u_{\xi}+\xi + i \rho}{\sqrt{2}}
		\end{equation}
		The physical mass square terms of the scalars are then given by 
		\begin{eqnarray}
		M^2_{h}&\simeq& 2\lambda_Hv^2+\frac{\lambda_{H \Phi_{12}}v^2_{12}}{2}+\frac{\lambda_{H \Phi_3}v^2_{3}}{2}+\frac{\lambda_{H \Phi_{B-L}}v^2_{B-L}}{2}+\frac{\mu u_{_{\Delta}}}{2}   \nonumber \\
		M^2_{\delta}& \simeq & -\frac{\mu v^2}{4\sqrt{2}u_{_\Delta}}-\frac{yv^2_{B-L}u_{\xi}}{8u_{_{\Delta}}}\nonumber \\
		M^2_{\xi}&\simeq& -\frac{yv^2_{B-L}u_{_{\Delta}}}{8u_{\xi}}\nonumber\\
		M^2_{\phi_{12}}&\simeq& 2\lambda_{\Phi_{12}} v^2_{12}+\frac{\lambda_{H \Phi_{12}}v^2}{2}+\frac{\lambda_{\Phi_{12} \Phi_3}v^2_3}{2}\nonumber\\&+&\frac{\lambda'_{\Phi_{12} \Phi_3}v^2_3}{2}+\frac{\lambda_{\Phi_{12} \Phi_{B-L}}v^2_{B-L}}{2}+\frac{\lambda'_{\Phi_{12} \Phi_{B-L}}v^2_{B-L}}{2}+\frac{Y v^2_{B-L}v_3}{8v_{12}} \nonumber \\
		M^2_{\phi_3}&\simeq& 2\lambda_{\Phi_3} v^2_{3}+\frac{\lambda_{H \Phi_3}v^2}{2}+\frac{\lambda_{\Phi_{12} \Phi_3}v^2_{12}}{2}\nonumber\\&+&\frac{\lambda'_{\Phi \Phi_3}v^2_{12}}{2}+\frac{\lambda_{\Phi_3 \Phi_{B-L}}v^2_{B-L}}{2}+\frac{\lambda'_{\Phi_3 \Phi_{B-L}}v^2_{B-L}}{2}+\frac{Y v^2_{B-L}v_{12}}{8v_{3}}\nonumber\\
		M^2_{\phi_{B-L}}&\simeq& 2\lambda_{\Phi_{B-L}}v^2_{B-L} + \frac{\lambda_{H \Phi_{B-L}}v^2}{2}+\frac{\lambda_{\Phi_{12} \Phi_{B-L}}v^2_{12}}{2}\nonumber\\&+&\frac{\lambda'_{\Phi_{12} \Phi_{B-L}}v^2_{12}}{2}+\frac{\lambda_{\Phi_3 \Phi_{B-L}}v^2_{3}}{2}+\frac{\lambda'_{\Phi_3 \Phi_{B-L}}v^2_{3}}{2}+\frac{Y v^2_{12}v_3}{2}
		\nonumber\\
		\end{eqnarray}
		where we have neglected the higher order terms involving $u_{\Delta}$ and $u_{\xi}$ as it is very small and no significant contribution comes from these terms.
		
		The $Z_{B-L}$ boson acquires mass through the vevs of $\Phi_{{B-L}}$, $\Phi_{12}$, $\Phi_3$ which are charged under $U(1)_{{B-L}}$ and is given by:
		\begin{equation}
		M_{Z_{B-L}}^{2}=g^2_{_{B-L}}(v^2_{{B-L}}+64v^2_{12}+100v^2_3).
		\end{equation}
		
		For simplicity as well as from interesting phenomenological perspective, we assume that the masses of $h$, $\phi_3$ and $\xi$ are of similar order in 
		sub-TeV range, while the masses of $\phi_{ B-L}$ and $\phi_{12}$ are in a few TeV scale. Thus the assumption for mass heirarchy among scalars is 
		$M_{\delta} >> M_{\phi_{_{B-L}}}, M_{\phi_{12}} > M_h, M_{\phi_3}, M_\xi$. Now we examine the mixing among the CP-even scalars $h,\phi_3$ and $\xi$. The 
		mixing between $h$ and $\xi$ is highly suppressed as we discussed in section \ref{neutrinomass} and hence negligible. Similarly $\xi$-$\Phi_{3}$ 
		mixing is of the order $\mathrm{O}(\frac{u_{\xi}}{v_{3}})$ and hence negligibly small. Hence in what follows we consider $H$ and $\Phi_3$ 
		mixing while discussing low energy phenomenology.

		Minimising the scalar potential \ref{scalar_potential} with respect to $\langle H \rangle =v $ and $ \langle \Phi_3 \rangle = v_3$, we obtain:
		\begin{equation}
		v_3=\sqrt{\frac{2\lambda_{H \Phi_3 }M^2_H - 4\lambda_H M^2_{\Phi_3}}{4\lambda_H \lambda_{\Phi_3}-\lambda^2_{H \Phi_3 }}}
		~~~{\rm and}~~~ v=\sqrt{\frac{2\lambda_{H \Phi_3 }M^2_{\Phi_3} - 4\lambda_{\Phi_3} M^2_{H}}{4\lambda_H \lambda_{\Phi_3}-\lambda^2_{H \Phi_3 }}}  .  
		\end{equation}
		
		The mass matrix of $h$ and $\phi_3$ can be written as:
		\begin{equation}
		\mathcal{M}^2(h,\phi_3)=
		\begin{pmatrix}
		\lambda_H v^2 & \lambda_{H \Phi_3}v v_3\\
		\lambda_{H \Phi_3}v v_3 & \lambda_{\Phi_3} v^2_3\\
		\end{pmatrix}.
		\end{equation}
		Here the mixing between $h$ and $\phi_3$ is governed by the coupling $\lambda_{H \Phi_3}$. The masses of the physical Higgses can be obtained by diagonalising the above mass matrix as:
		\begin{eqnarray}
		M^2_{h_1} &=& \frac{1}{2}[(\lambda_{H}v^2+\lambda_{\Phi_3}v^2_3)- \sqrt{(\lambda_{\Phi_3}v^2_3 - \lambda_{H}v^2)^2 + 4 (\lambda_{H \Phi_3}v v_3)^2  }] \nonumber\\
		M^2_{h_2} &=& \frac{1}{2}[(\lambda_{H}v^2+\lambda_{\Phi_3}v^2_3)+ \sqrt{(\lambda_{\Phi_3}v^2_3 - \lambda_{H}v^2)^2 + 4 (\lambda_{H \Phi_3}v v_3)^2  }]
		\end{eqnarray}
		where $h_{1}$ and $h_{2}$ can be realised as SM-like Higgs and the second Higgs, respectively. Thus the mass eigen states of these scalars 
		can be given as:
		\begin{eqnarray}
		h_1 &=& \cos\beta ~h + \sin \beta ~\phi_3 \nonumber\\
		h_2 &=& -\sin\beta ~h + \cos\beta ~\phi_3 \,,
		\end{eqnarray}
		where 
		\begin{equation}
		\tan 2 \beta= \Big(\frac{2 \lambda_{H \Phi_3}v v_3}{\lambda_{\Phi_3} v^2_3- \lambda_{H}v^2}\Big).
		\end{equation}
		Thus it is evident that the mixing angle $\beta$ vanishes if $\lambda_{H \Phi_3}\rightarrow 0 $ or if $v_3>>v$.
		
		
		\section{Dark Matter}
		At a TeV scale, the $U(1)_{\rm B-L}$ gauge symmetry breaks down to a remnant $Z_2$ by the vev of $\Phi_{12}$, $\Phi_{\rm B-L}$ 
		and $\Phi_3$. We assume that the right-handed neutrinos are odd under the $Z_2$ symmetry, while all other particles even. As 
		a result the lightest right-handed neutrino is a viable candidate of dark matter. 
		
		\subsection{Right Handed Neutrinos and Their Interactions}\label{RHN_interactions}
		
		From Eqn. (\ref{full_Lagr}), the mass matrix of right handed neutrinos in the effective theory can be given as: 
		\begin{equation}
		-\mathscr{L}_{\nu_R}^{mass}= \frac{1}{2}
		\begin{pmatrix}
		\overline{(\nu_{_{R1}})^c} &
		\overline{(\nu_{_{R2}})^c} & 
		\overline{(\nu_{_{R3}})^c} \\
		\end{pmatrix}
		\mathcal{M}
		\begin{pmatrix}
		\nu_{_{R1}} \\
		\nu_{_{R2}} \\
		\nu_{_{R3}} \\ 
		\end{pmatrix}
		\end{equation}
		where
		\begin{equation}
		\mathcal{M}= 
		\begin{pmatrix}
		Y_{11}v_{_{12}} && Y_{12}v_{_{12}} && Y_{13}v_{_{B-L}}\\
		Y_{12}v_{_{12}} && Y_{22}v_{_{12}} && Y_{23}v_{_{B-L}}\\
		Y_{13}v_{_{B-L}} && Y_{23}v_{_{B-L}} && Y_{33}v_{3}\\ 
		\end{pmatrix} = 
		\begin{pmatrix}  
		[M_{12}] &&  [M'] \\
		[M']^T   &&  M_3  \\ 
		\end{pmatrix}.
		\end{equation}
		Here $M_{12}, M', M_3$ are:
		\begin{center}
			$M_{12}= 
			\begin{pmatrix}
			Y_{11} v_{_{12}} && Y_{12} v_{_{12}}\\
			Y_{12} v_{_{12}} && Y_{22} v_{_{12}}\\
			\end{pmatrix}$
			, 
			$M'=
			\begin{pmatrix} 
			Y_{13}v_{_{B-L}}\\
			Y_{23}v_{_{B-L}}\\
			\end{pmatrix}$
			, 
			$M_3=
			\begin{pmatrix} 
			Y_{33}v_{3}
			\end{pmatrix}$.
		\end{center}
		
		
		The above right handed neutrino Majorana mass matrix $\mathcal{M}$ can be block-diagonalised using a rotation matrix of the form :
		\begin{equation}
		\mathcal{R_\theta}= \begin{pmatrix} \cos\theta \begin{bmatrix} 1 & 0\\
		0 & 1\end{bmatrix} && \sin\theta \begin{bmatrix} 1 \\ 1 \end{bmatrix} \\ -\sin\theta \begin{bmatrix} 1 && 1 \end{bmatrix} &&\cos\theta \end{pmatrix} = \begin{pmatrix}\cos\theta & 0 & \sin\theta \\
		0 & \cos\theta & \sin\theta \\                          -\sin\theta & -\sin\theta & \cos\theta\\ 
		\end{pmatrix} 
		\end{equation}
		where it can be checked that this transformation is orthogonal in the linear over $\theta$ approximation as, 
		\begin{equation}
		(\mathcal{R_\theta})^T \mathcal{R_\theta} = \begin{pmatrix}
		1 & {\sin^2\theta } & 0 \\
		\sin^2\theta & 1 & 0 \\
		0 & 0 & 1+\sin^2\theta \\
		\end{pmatrix} \simeq I_{3}.
		\end{equation}
		
		Here $\theta$ is essentially the mixing between $M_{12}$ and $M_{3}$. This mixing angle can be expressed as:
		
		\begin{equation}
		\sin2\theta \simeq \frac{Y_{13}v_{_{B-L}}}{Y_{11}v_{12}+ Y_{12}v_{12}-Y_{33}v_{3}} \simeq \frac{Y_{23}v_{_{B-L}}}{Y_{11}v_{12}+ Y_{12}v_{12}-Y_{33}v_{3}}.
		\label{sin2theta}
		\end{equation}

		Then we can completely diagonalise the block-diagonalised matrix $\mathcal{M}$ further by performing another orthogonal transformation using the transformation matrix
		
		\begin{equation}
		\mathcal{R_\alpha} = 
		\begin{pmatrix}
		\cos\alpha  & \sin\alpha & 0 \\
		-\sin\alpha & \cos\alpha & 0 \\                          
		0      &     0      & 1 \\ 
		\end{pmatrix}\,,
		\end{equation}
		where the mixing angle $\alpha$ is given by 
		\begin{equation}
		\tan2\alpha \simeq \frac{Y_{12}v_{12}}{Y_{11}v_{12}-Y_{22}v_{12}}.
		\end{equation}
		
		Thus the mass eigen states of the right handed neutrinos can be written as:
		\begin{equation}
		\begin{pmatrix}
		N_{1R}\\
		N_{2R}\\
		N_{3R}\\
		\end{pmatrix} 
		={\begin{pmatrix}
			\cos\alpha & \sin\alpha & 0 \\
			-\sin\alpha & \cos\alpha & 0 \\                          
			0 & 0 & 1\\ 
			\end{pmatrix}}
		\begin{pmatrix}
		\cos\theta   &      0       &   \sin\theta \\
		0        &  \cos\theta  &   \sin\theta \\
		-\sin\theta  & -\sin\theta  &   \cos\theta \\ 
		\end{pmatrix}
		\begin{pmatrix}
		\nu_{_{R1}} \\
		\nu_{_{R2}} \\
		\nu_{_{R3}} \\
		\end{pmatrix}
		\end{equation}
		%
		%
		Here it is worth noting that the matrix $\mathcal{R}=\mathcal{R_\alpha} \mathcal{R_\theta}$ is also orthogonal (${\it i.e.,}~ \mathcal{R}^T \mathcal{R}= I$) in the linear over $\theta$ approximation. 
		
		For simplicity we assume that there is strong hierarchy between $\nu_{R3}$ and $\nu_{R1}$, $\nu_{R2}$. This implies that 
		$\sin \theta \to 0$ in Eq.\,\ref{sin2theta}. In fact, this can be achieved if we assume $Y_{13} v_{\rm B-L}, Y_{23} v_{\rm B-L} 
		<< (Y_{11}+Y_{12})v_{12}$. In this limit, $\nu_{_{R3}}$ completely decouples from $\nu_{_{R1}}$ and $\nu_{_{R2}}$. As a result 
		the diagonalisation of above mass matrix gives the mass eigen values corresponding to the states $N_{1R}$,$N_{2R}$ and $N_{3R}$ as
		\begin{eqnarray}
		M_{1,2}  &=& \frac{1}{2}[({Y_{11}v_{12}+Y_{22}v_{12}}) \pm \sqrt{(Y_{11}v_{12}-Y_{22}v_{12})^2+4(Y_{12}v_{12})^2}] \nonumber\\
		M_{3} &=& Y_{33}\, v_3.
		\end{eqnarray}

		
		\subsubsection*{Interactions}
		The interaction terms of the right handed neutrinos with $Z_{B-L}$ in the mass eigen basis can be written as :
		\begin{align}
		\mathcal{L}_{Z_{B-L}}&=g_{_{_{B-L}}}\bigg[\{ (4\cos^2 \theta -5 \sin^2 \theta)(1+ \sin 2\alpha)\} \overline{N_{1R}}\gamma^\mu N_{1R}+ \{  16 \sin^2 \theta - 5 \cos^2 \theta \} \overline{N_{3R}}\gamma^\mu N_{3R} \nonumber\\
		&  +  \{ (4\cos^2 \theta -5 \sin^2 \theta)(1-\sin 2\alpha)\} \overline{N_{2R}}\gamma^\mu N_{2R} \nonumber\\
		&  +  \cos2\alpha \{ 4\cos^2 \theta -5 \sin^2 \theta \}\Big(\overline{N_{1R}}\gamma^\mu N_{2R} + \overline{N_{2R}}\gamma^\mu N_{1R}\Big) \nonumber\\
		&  - \frac{13}{2}\sin2\theta \{\cos \alpha + \sin \alpha \}\Big(\overline{N_{1R}}\gamma^\mu N_{3R} + \overline{N_{3R}}\gamma^\mu N_{1R}\Big) \nonumber\\
		&  - \frac{13}{2}\sin2\theta \{\cos \alpha - \sin \alpha \}\Big(\overline{N_{2R}}\gamma^\mu N_{3R} + \overline{N_{3R}}\gamma^\mu N_{2R}\Big) \bigg] (Z_{B-L})_\mu.
		\end{align}

		The Yukawa interactions in the mass basis can be written as:
		\begin{align}
		Y_{ab} \overline{(\nu_{_{Ra}})^c} \Phi \nu_{_{Rb}}&=Y_{11} \overline{(\nu_{_{R1}})^c} \Phi \nu_{_{R1}}+Y^{22} \overline{(\nu_{_{R2}})^c} \Phi \nu_{_{R2}}+Y^{12} \overline{(\nu_{_{R1}})^c} \Phi \nu_{_{Rb}}+Y^{12} \overline{(\nu_{_{R2}})^c} \Phi \nu_{_{R1}}\nonumber\\
		& =  Y\bigg[\cos^2\theta(1+\sin2\alpha)\overline{(N_{1R})^c} \phi N_{1R}+\cos^2\theta(1-\sin2\alpha)\overline{(N_{2R})^c} \phi N_{2R}\nonumber\\
		& +  4\sin^2\theta \overline{(N_{3R})^c} \phi N_{3R}+\cos^2\theta \cos2\alpha \Big( \overline{(N_{1R})^c} \phi N_{2R} + \overline{(N_{2R})^c} \phi N_{1R} \Big) \nonumber\\
		& -  \sin2\theta(\cos\alpha+\sin\alpha) \Big( \overline{(N_{1R})^c} \phi N_{3R} + \overline{(N_{3R})^c} \phi N_{1R}\Big)\nonumber\\
		& -  \sin2\theta(\cos\alpha-\sin\alpha)\Big( \overline{(N_{2R})^c} \phi N_{3R} + \overline{(N_{3R})^c} \phi N_{2R}\Big) \bigg]
		\end{align}
		Similarly 
		\begin{align}
		Y_{a3} \overline{(\nu_{_{Ra}})^c} \Phi_{{B-L}} \nu_{_{R3}}&= Y\Big[ \overline{(\nu_{_{R1}})^c} \Phi_{{B-L}} \nu_{_{R3}}+ \overline{(\nu_{_{R2}})^c} \Phi_{{B-L}} \nu_{_{R3}} + h.c\Big]\nonumber\\
		& =Y\bigg[\sin2\theta(1+\sin2\alpha)\overline{(N_{1R})^c} \phi_{{{B-L}}} N_{1R} + \sin2\theta(1-\sin2\alpha)\overline{(N_{2R})^c} \phi_{{{B-L}}} N_{2R}\nonumber\\&-2\sin2\theta \overline{(N_{3R})^c} \phi_{{{B-L}}} N_{3R}\nonumber\\& +\frac{1}{2}\sin2\theta(\cos2\alpha+\sin2\alpha)
		\Big(\overline{(N_{1R})^c} \phi_{{{B-L}}} N_{2R} + \overline{(N_{2R})^c} \phi_{{{B-L}}} N_{1R}\Big) \nonumber\\&+ (\cos^2\theta-2\sin^2\theta)(\cos\alpha-\sin\alpha)\Big( \overline{(N_{2R})^c} \phi_{{{B-L}}} N_{3R} + \overline{(N_{3R})^c} \phi_{{{B-L}}} N_{2R}\Big) \nonumber\\&+(\cos^2\theta-2\sin^2\theta)(\cos\alpha+\sin\alpha)\Big( \overline{(N_{2R})^c} \phi_{{{B-L}}} N_{3R} + \overline{(N_{3R})^c} \phi_{{{B-L}}} N_{2R}\Big)  \bigg].
		\end{align}
		
		and
		\begin{align}
		Y_{33} \overline{(\nu_{_{R3}})^c} \Phi_3 \nu_{_{R3}} &= Y_{33}\bigg[\sin^2\theta(1+\sin2\alpha)\overline{(N_{1R})^c} \phi_3 N_{1R}+\sin^2\theta(1-\sin2\alpha)\overline{(N_{2R})^c} \phi_3 N_{2R}\nonumber\\
		& +  \cos^2\theta \overline{(N_{3R})^c} \phi_3 N_{3R}+\sin^2\theta \cos2\alpha \Big( \overline{(N_{1R})^c} \phi_3 N_{2R} + \overline{(N_{2R})^c} \phi_3 N_{1R}\Big) \nonumber\\
		& +  \frac{1}{2}\sin2\theta(\cos\alpha+\sin\alpha)\Big( \overline{(N_{1R})^c} \phi_3 N_{3R} + \overline{(N_{3R})^c} \phi_3 N_{1R}\Big) \nonumber\\
		& +  \frac{1}{2}\sin2\theta(\cos\alpha-\sin\alpha)\Big( \overline{(N_{2R})^c} \phi_3 N_{3R} + \overline{(N_{3R})^c} \phi_3 N_{2R}\Big) \bigg].
		\label{nu_R_phi3_int}
		\end{align}

		But as $\phi_3$ mixes with SM Higgs $h$, then the Eq.\,\ref{nu_R_phi3_int} can be written in terms of the scalar mass eigen states as:

		\begin{align}
		Y_{33} \overline{(\nu_{_{R3}})^c} \Phi_3 \nu_{_{R3}} &= Y_{33}\sin\beta \bigg[\sin^2\theta(1+\sin2\alpha)\overline{(N_{1R})^c} h_1 N_{1R}+\sin^2\theta(1-\sin2\alpha)\overline{(N_{2R})^c} h_1 N_{2R}\nonumber\\
		& +  \cos^2\theta \overline{(N_{3R})^c} h_1 N_{3R}+\sin^2\theta \cos2\alpha \Big( \overline{(N_{1R})^c} h_1 N_{2R} + \overline{(N_{2R})^c} h_1 N_{1R}\Big) \nonumber\\
		& +  \frac{1}{2}\sin2\theta(\cos\alpha+\sin\alpha)\Big( \overline{(N_{1R})^c} h_1 N_{3R} + \overline{(N_{3R})^c} h_1 N_{1R}\Big) \nonumber\\
		& +  \frac{1}{2}\sin2\theta(\cos\alpha-\sin\alpha)\Big( \overline{(N_{2R})^c} h_1 N_{3R} + \overline{(N_{3R})^c} h_1 N_{2R}\Big) \bigg] \nonumber\\
		& +  Y^{33}\cos\beta \bigg[\sin^2\theta(1+\sin2\alpha)\overline{(N_{1R})^c} h_2 N_{1R}+\sin^2\theta(1-\sin2\alpha)\overline{(N_{2R})^c} h_2 N_{2R}\nonumber\\
		& +  \cos^2\theta \overline{(N_{3R})^c} h_2 N_{3R}+\sin^2\theta \cos2\alpha \Big( \overline{(N_{1R})^c} h_2 N_{2R} + \overline{(N_{2R})^c} h_2 N_{1R}\Big) \nonumber\\
		& +  \frac{1}{2}\sin2\theta(\cos\alpha+\sin\alpha)\Big( \overline{(N_{1R})^c} h_2 N_{3R} + \overline{(N_{3R})^c} h_2 N_{1R}\Big) \nonumber\\
		& +  \frac{1}{2}\sin2\theta(\cos\alpha-\sin\alpha)\Big( \overline{(N_{2R})^c} h_2 N_{3R} + \overline{(N_{3R})^c} h_2 N_{2R}\Big) \bigg].
		\label{nu3_nu3_phi_co}
		\end{align}
	\begin{figure}
		\begin{center}		
		\includegraphics[scale=0.9]{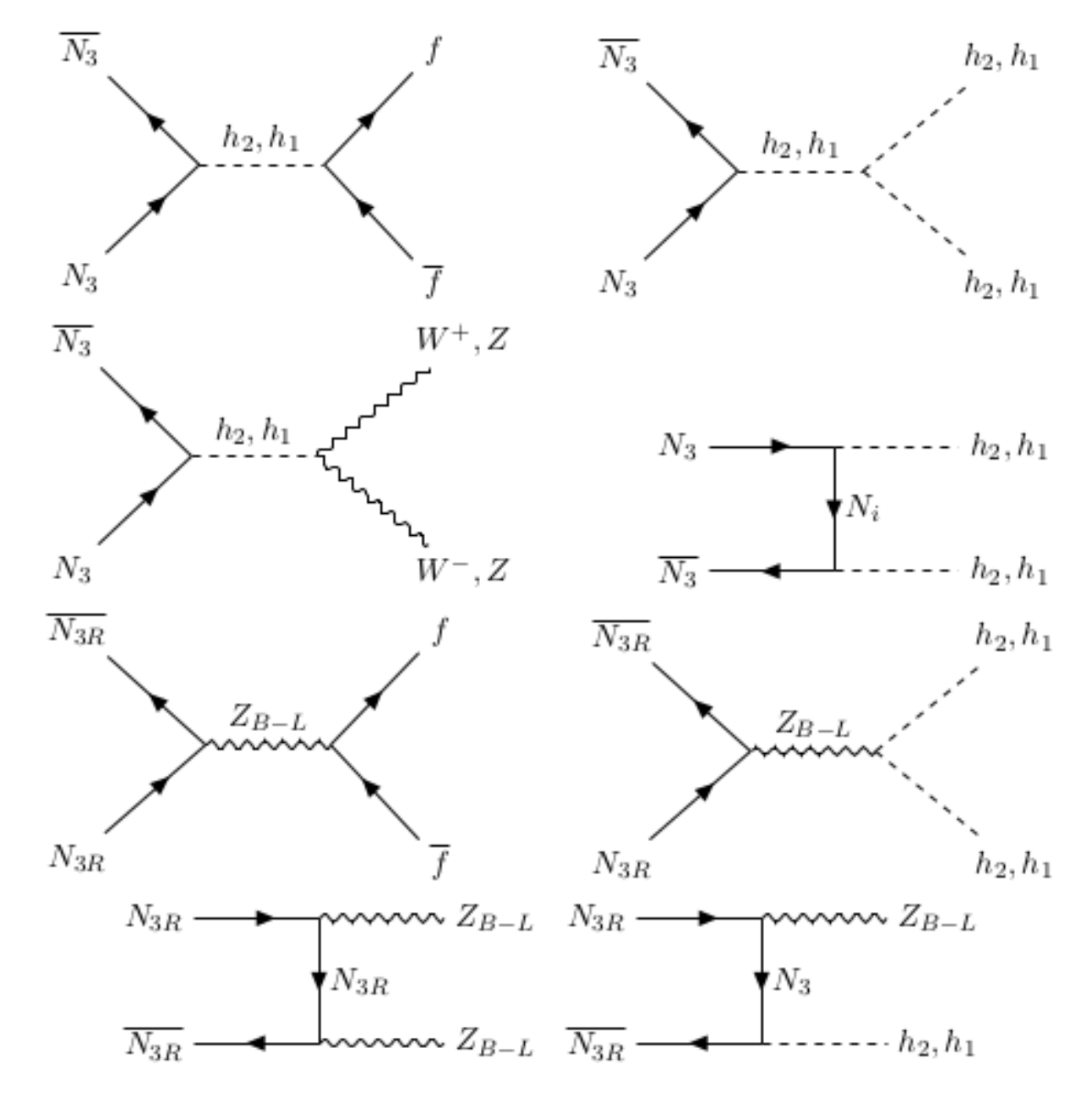}
	\end{center}
			\caption{DM annihilation channels.}
			\label{DM_anni}
		\end{figure}
		\subsection{Relic Abundance of DM}
		As discussed above, the lightest right-handed neutrino is stable due to a remnant $Z_2$ symmetry. Without loss of generality we assume 
		$N_{3}=(N_{3R}+ N_{3R}^c)/\sqrt{2}$ is the lightest right handed neutrino and acts as a candidate of DM.  
		
		
		The relic abundance of DM ($N_3$) can be achieved from the annihilation channels of $N_{3}$ to the SM particles. The various processes which are contributing to the DM relic abundance via the annihilation channels can be seen in Fig.\,\ref{DM_anni}.
		
\begin{figure}[h!]
			\centering
			\includegraphics[width = 100mm]{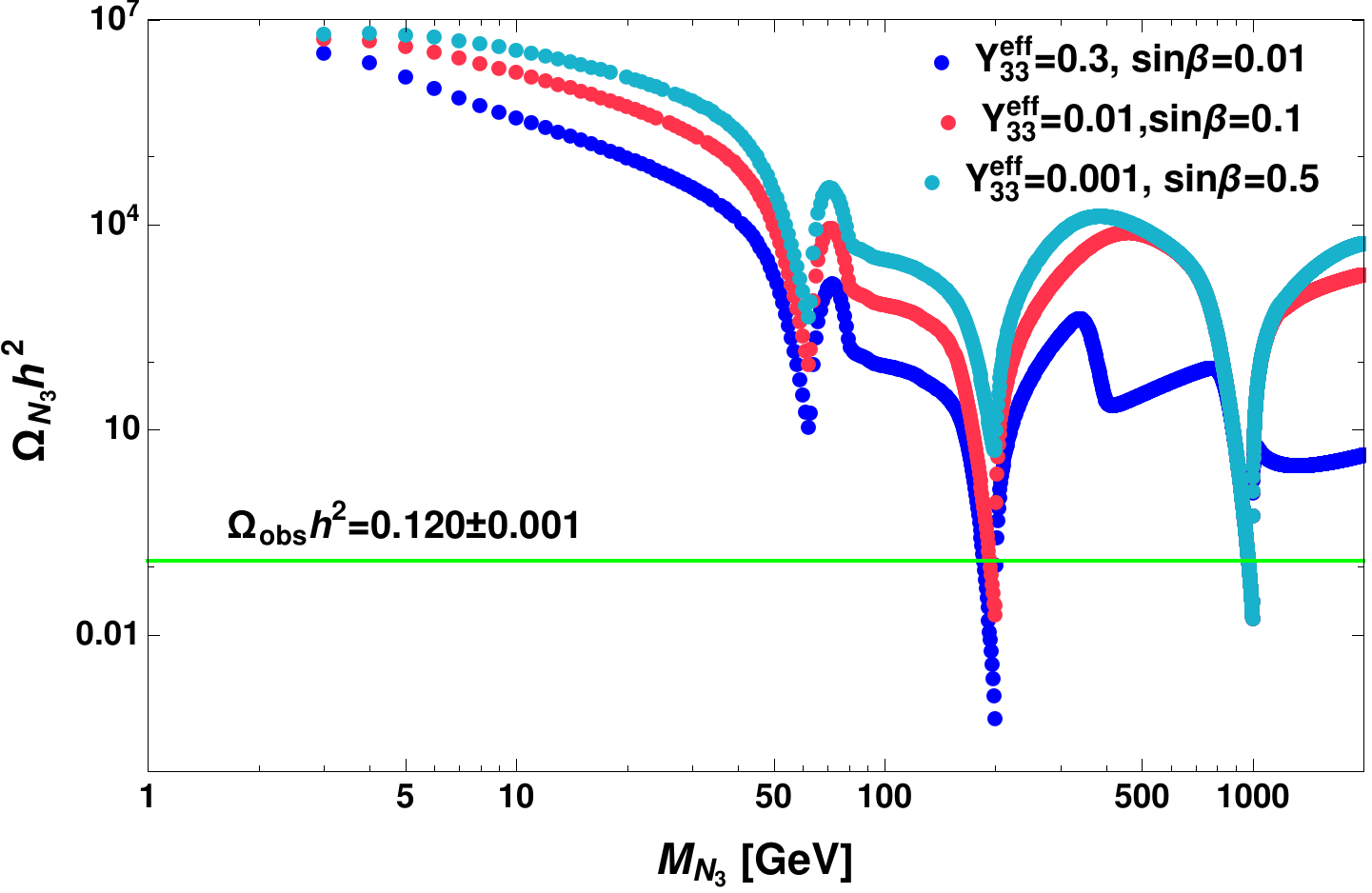}
			\caption{\footnotesize{Relic density of DM as a function of its mass. Contribution to the relic density of DM is only through its annihilation channels. The Green horizontal line shows the observed relic density of DM \cite{Aghanim:2018eyx}.}}
			\label{DM_relic}
		\end{figure}

\begin{figure}[h!]
			\centering
			\includegraphics[width = 100mm]{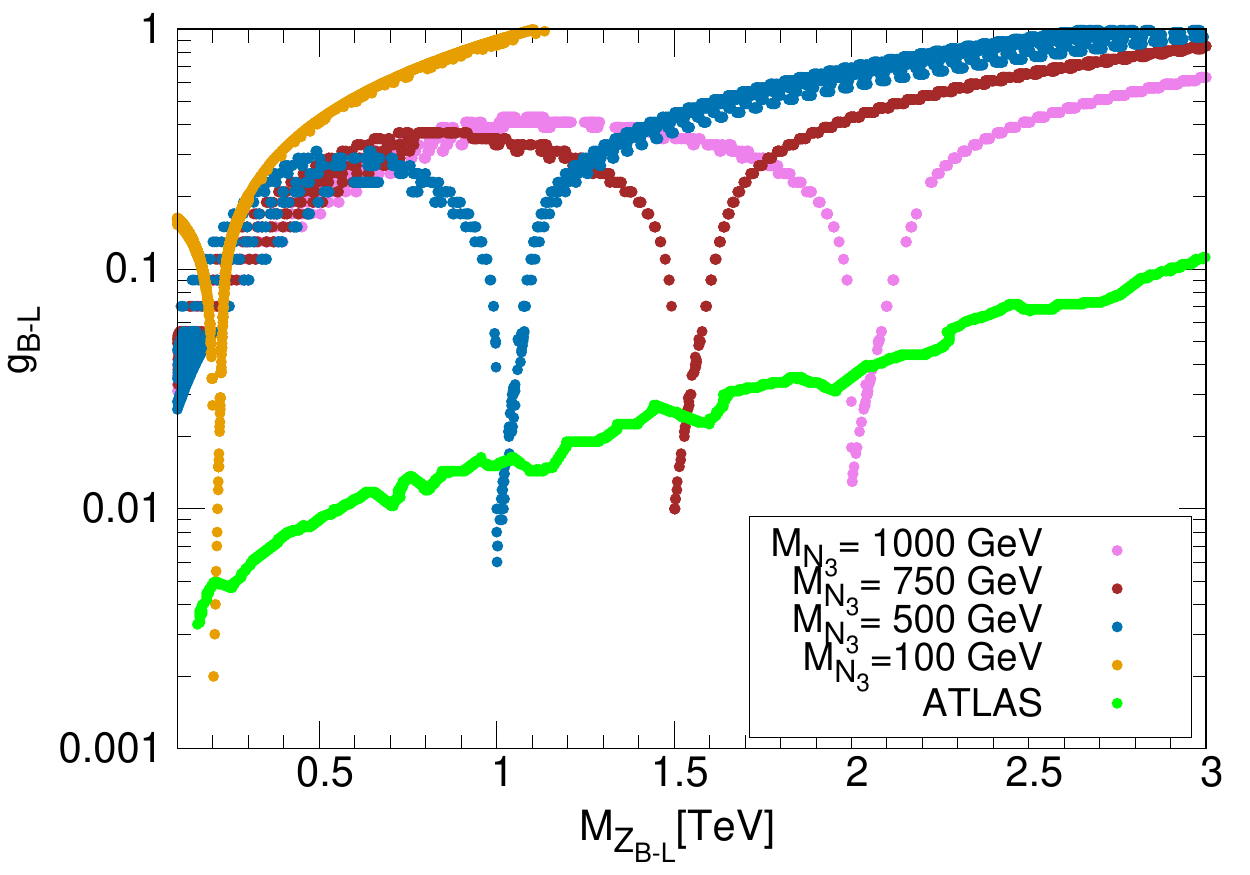}
			\caption{\footnotesize{ Points allowed by the relic density constraint in the plane of $M_{Z_{B-L}}$ vs $g_{B-L}$. The dips corresponds to the resonance for the mass of $Z_{B-L}$, ${\it i.e.,}$ $M_{Z_{B-L}}$=200(Orange), 1000(Blue), 1500(Maroon), 2000(Pink) GeV, corresponding to the DM mass 100, 500, 750 and 1000 GeV respectively. The Geen line shows the ATLAS constraint on effectibe coupling strength $g_{_{B-L}}$ as a function of gauge boson mass~\cite{Aaboud:2017buh}. Here the effective coupling $Y_{33}^{eff}$ and $\sin \beta$ are taken to be random numbers between $10^{-3}-1$.}}
			\label{DM_mzbl_gbl}
		\end{figure}

\begin{figure}[h!]
			\centering
			\includegraphics[width = 100mm]{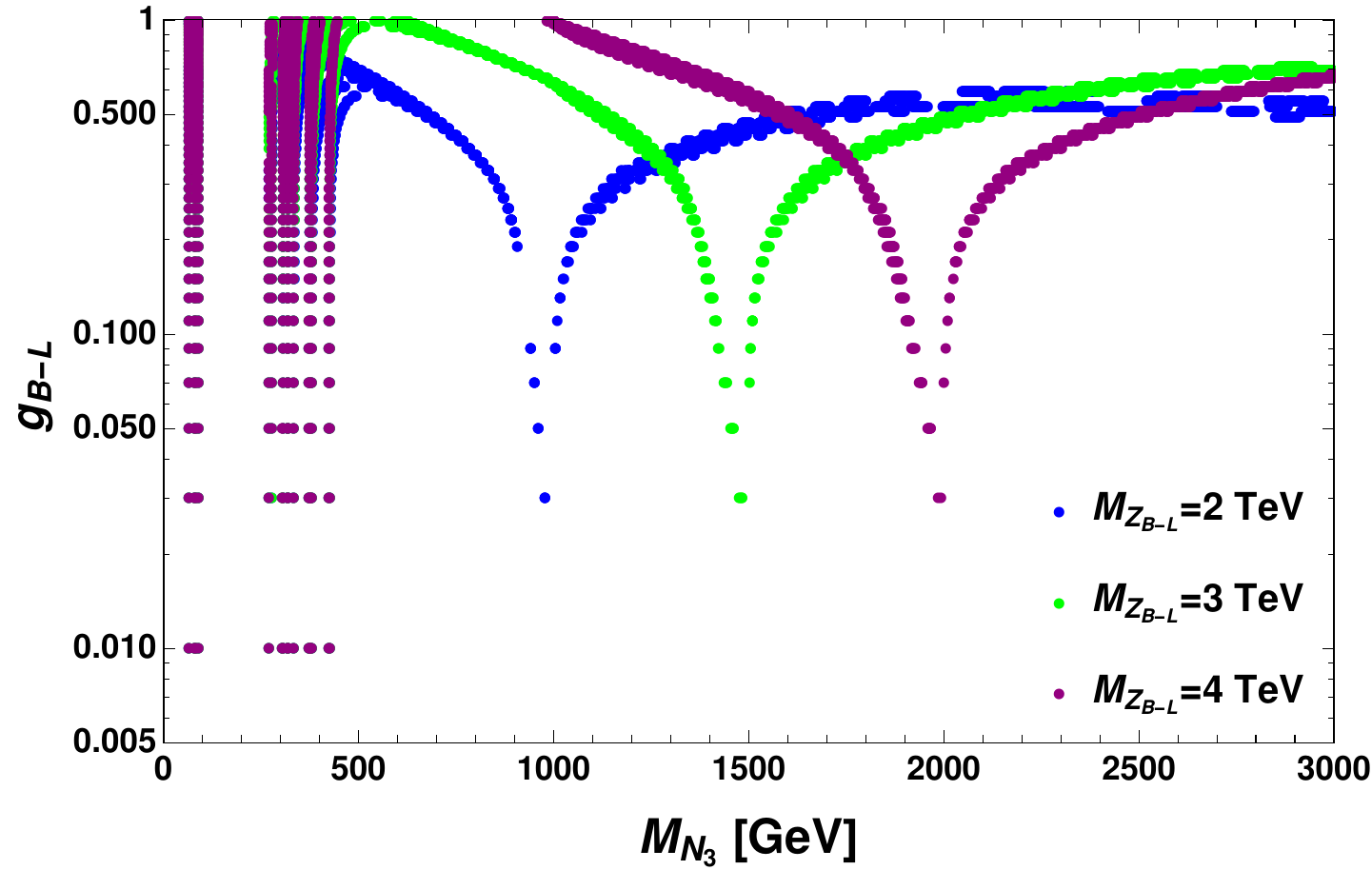}
			\caption{\footnotesize{Points allowed by the relic density constraint in the plane of $M_{N_{3}}$ vs $g_{B-L}$. The dips correspond to the resonance corresponding to the mass of $Z_{B-L}$, ${\it i.e.,}$ $M_{Z_{B-L}}$=2000(Blue), 3000(Green), 4000(Purple) GeV. The effective coupling $Y_{33}^{eff}$ and $\sin \beta$ are taken to be random numbers between $10^{-3}-1$.}}
			\label{DM_mn3_gbl}
		\end{figure}

		The relic density of DM is calculated using micrOMEGAs~\cite{Belanger:2008sj}. The masses for the heavier $Z_2$ odd particles are chosen to be 1.5 TeV and 2 TeV.  In Fig.\,\ref{DM_relic}, we show the relic density of DM as a function of DM mass. The resonance regions are corresponding to the SM-like Higgs $h_{1}$, second Higgs $h_{2}$ and gauge boson $Z_{B-L}$. The masses of second Higgs $M_{h_{2}}$ and the gauge boson $M_{Z_{B-L}}$ are taken to be 400 GeV and 2000 GeV respectively and the gauge coupling is taken to be $g_{_{B-L}}=0.035$, where these value are in agreement with the collider constraints~\cite{Aaboud:2017buh}. In Fig.\,\ref{DM_relic}, an arbitrary choice of set ($Y_{33}^{eff},~\sin \beta)$ is taken $(0.3, 0.01), (0.01, 0.1), (0.001, 0.5)$. One can find that the relic abundance criteria of DM is typically  satisfied at the resonance regions only. The effective coupling $Y_{33}^{eff}$ of the vertex $\overline{(\nu_{R3})^{c}} h_2 \nu_{R3}$ and $\sin \beta$ are taken to be random numbers between $10^{-3}-1$ and the points which are allowed for $Y_{33}^{eff}$ and $\sin \beta$ by the relic density constraint are shown in Fig.\,\ref{DM_hee_SG} (Orange points). We also show the allowed parameter space in the $g_{_{B-L}} - M_{Z_{B-L}}$ plane which are allowed by the relic density constraints for three different values of DM mass in Fig~\ref{DM_mzbl_gbl}. Fig~\ref{DM_mn3_gbl} shows the plots of $g_{_{B-L}}$ vs $M_{N_3}$ which satisfies the observed relic density for $M_{Z_{B-L}}=$ 2 TeV, 3 TeV and 4 TeV respectively.  
		
		\vspace{0.5cm}
		
		\hspace{1cm}
		\subsection{Direct Detection}
\begin{figure}[h!]
			\centering
			\includegraphics[width = 40mm]{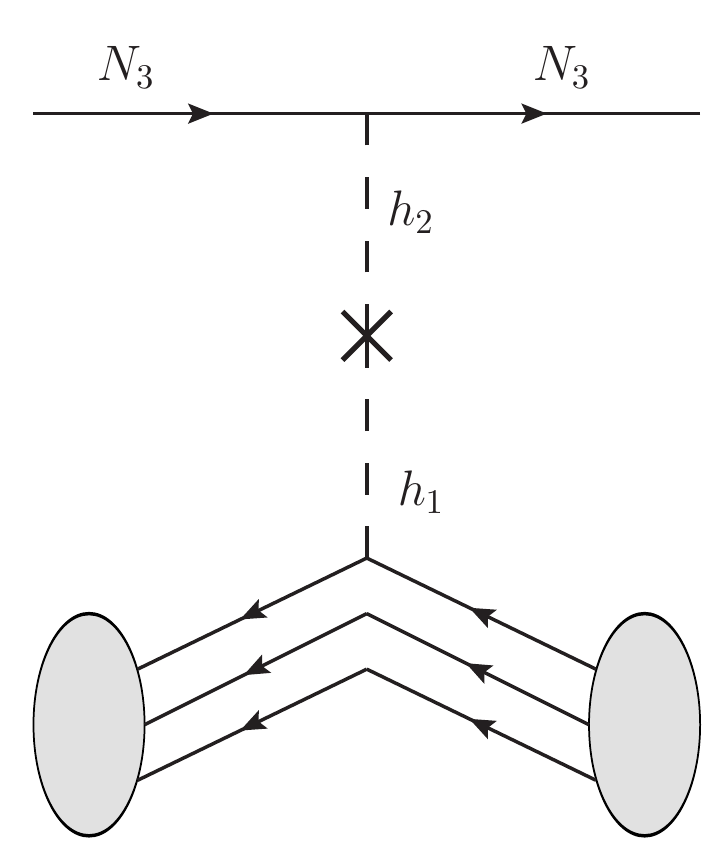}
			\caption{\footnotesize{The spin-independent scattering cross-section of DM-nucleon via Higgs portal.}}
			\label{DM_diag}
		\end{figure}

		The spin-independent scattering of DM is possible through $\phi_{3}-h$ mixing, where DM particles can scatter off the target nuclei which are located at terrestrial laboratories. The corresponding Feynmann diagram can be seen in Fig.\,\ref{DM_diag}. The spin-independent elastic scattering cross-section of DM per nucleon can be expressed as
		\begin{equation}
		\sigma_{SI}^{h_{1}h_{2}} = \frac{\mu_{r}^{2}}{\pi A^{2}} \left[ Z f_{p} + (A-Z) f_{n} \right]^{2}
		\label{DD_cs}
		\end{equation}
		
		where $\mu_{r}=M_{_{N_{3}}} m_{n}/(M_{_{N_{3}}} + m_{n})$ is the reduced mass, where $m_{n}$ is the nucleon (proton or neutron) mass.
		The A and Z are the mass and atomic number of the target nucleus, respectively. The $f_{p}$ and $f_{n}$ are the interaction strengths of proton and neutron with DM, respectively and they can be given as,
		\begin{equation}
		f_{p,n}=\sum\limits_{q=u,d,s} f_{T_{q}}^{p,n} \alpha_{q}\frac{m_{p,n}}{m_{q}} + \frac{2}{27} f_{TG}^{p,n}\sum\limits_{q=c,t,b}\alpha_{q} 
		\frac{m_{p,n}}{m_{q}}\,,
		\label{fpn}
		\end{equation}
		where 
		\begin{equation}
		\alpha_{q} = \frac{Y_{33}^{eff} \sin2\beta}{2\sqrt{2}} \left( \frac{m_{q}}{v}\right) \left[\frac{1}{M_{h_{2}}^{2}}-\frac{1}{M_{h_{1}}^{2}}\right] \,.
		\label{DD4}
		\end{equation}
		In the above Eq.\,\ref{fpn}, the values of $f_{T_{q}}^{p,n}$ can be found in~\cite{Ellis:2000ds}. 
		
		Using Eq.~\ref{fpn} and \ref{DD4}, the spin-independent cross-section in Eq.\,\ref{DD_cs}, can be re-expressed as:
		\begin{eqnarray}
		\sigma_{\rm SI}^{h_{1}h_{2}} &=& \frac{{\mu_{r}}^{2}}{\pi A^{2}} \left(\frac{Y_{33}^{eff} \sin 2 \beta}{2\sqrt{2}} \right)^{2}  \left[ \frac{1}{M_{h_{2}^{2}}}-\frac{1}{M_{h_{1}^{2}}} \right]^{2} \nonumber \\
		& \times & \left[ Z \left(\frac{m_{p}}{v}\right) \left(f_{Tu}^{p}+f_{Td}^{p}+f_{Ts}^{p}+\frac{2}{9}f_{TG}^{p} \right) + (A-Z) \left(\frac{m_{n}}{v}\right) \left(f_{Tu}^{n}+f_{Td}^{n}+f_{Ts}^{n}+\frac{2}{9}f_{TG}^{n}\right) \right]^{2}. \nonumber\\
		\label{SI_cross}
		\end{eqnarray}
		
\begin{figure}[h!]
			\centering
			\includegraphics[width = 100mm]{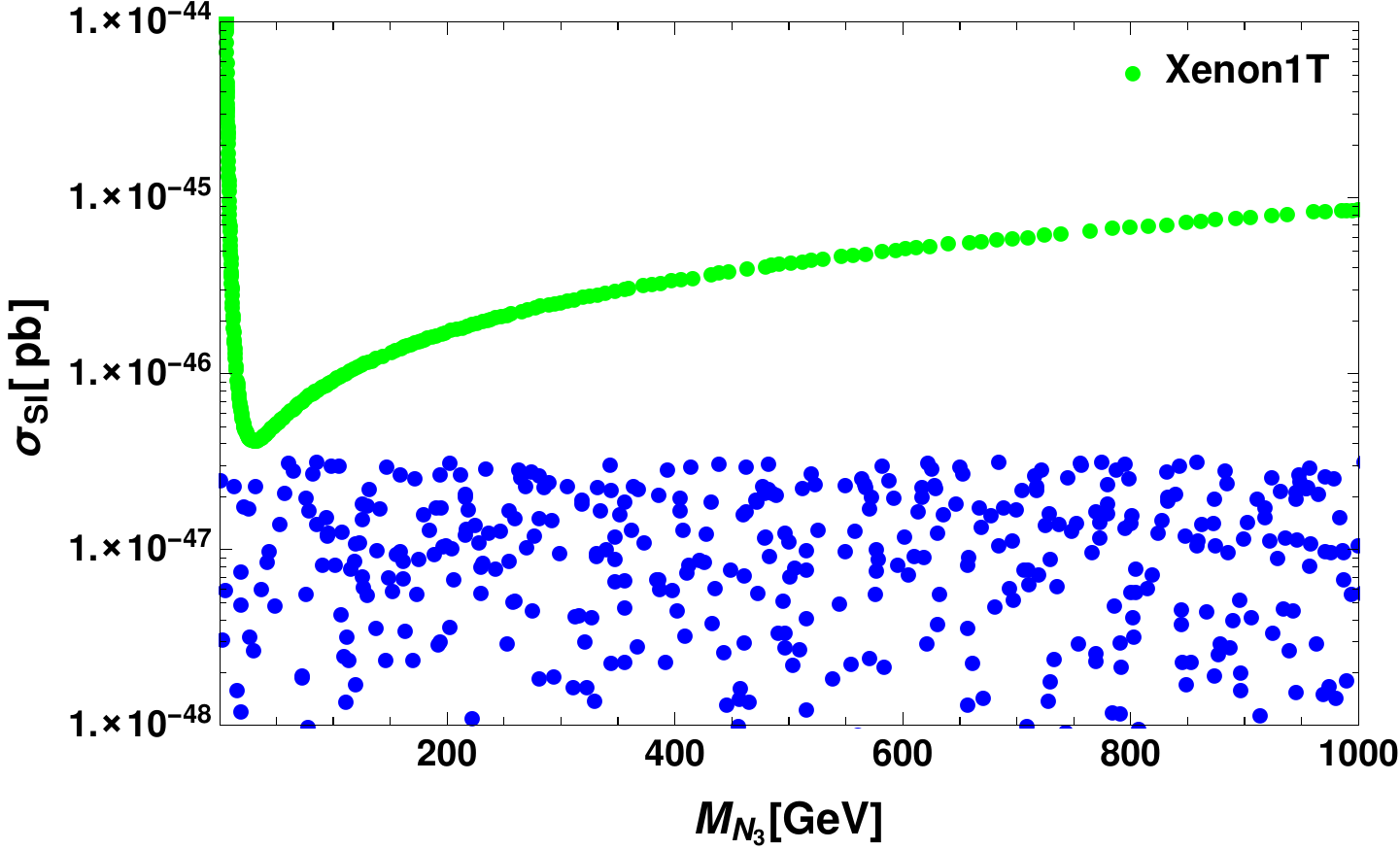}
			\caption{\footnotesize{The spin-independent DM-nucleon cross-section (Blue dots) which is not yet ruled out by XENON1T bound (Green dots).}}
			\label{DM_DD}
		\end{figure}

\begin{figure}[h!]
			\centering
			\includegraphics[width = 100mm]{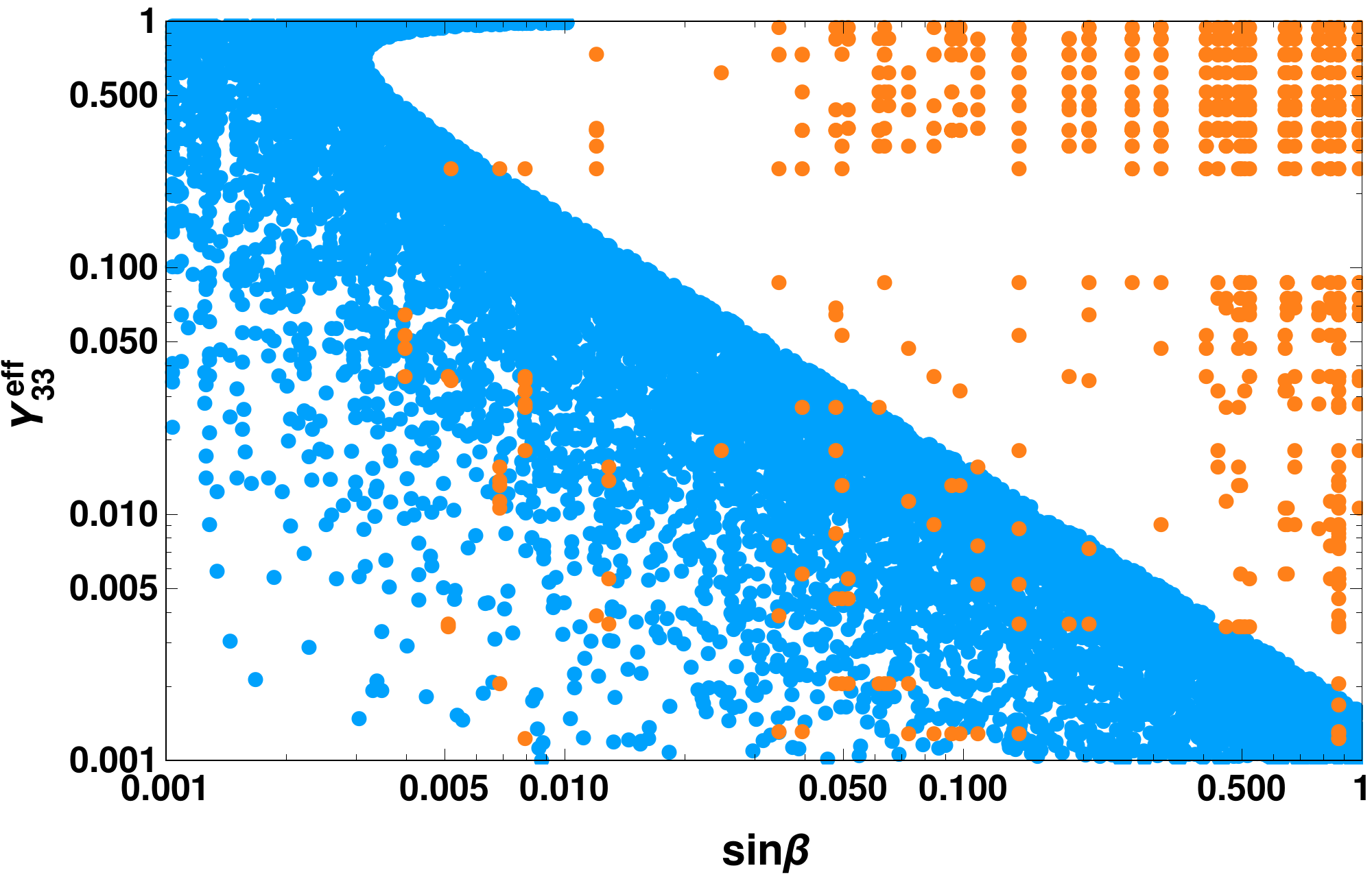}
			\caption{\footnotesize{The parameter region allowed by relic density and XENON1T bounds. The Orange points are allowed by the relic density constraint. The Sky blue points allowed by the XENON1T bound. }}
			\label{DM_hee_SG}
		\end{figure}

		The more stringent bound on direct detection come from the XENON1T\,\cite{Aprile:2018dbl}, where it rules out spin-independent cross-section down to $\sigma_{_{\rm SI}} \approx 10^{-47} {\rm cm}^{2}$. The Fig.\,\ref{DM_DD}, show the parameter space which is allowed by the XENON1T bound. The only free parameters in Eq.\,\ref{SI_cross} are $\lambda_{33}^{eff}$ and $\sin \beta$ These values are varied randomly between $10^{-3}-1$. The parameter space allowed by $\lambda_{33}^{eff}$ and $\sin \beta$ are shown in Fig.\,\ref{DM_hee_SG}, with Sky blue points, which keeps the spin-independent DM-nucleon cross-section below the XENON1T bound. The Orange points are allowed by the  relic density bound. The overlapped region are the points which are allowed by the both relic density and XENON1T bounds.
		
		\section{Collider Signatures}
		As the model is extended with $U(1)_{B-L}$ gauge group and all the fermions as well as most of the scalars like $\xi$,$\Phi_{_{B-L}}$,$\Phi_{12}$,$\Phi_3$  are charged under $B-L$ gauge interaction, so we can encounter interesting collider signatures predicted by this model at LHC or FCC~\cite{Majee:2010ar,Emam:2007dy,Melfo:2011nx}
		
		If $M_{Z_{B-L}} <$ $M_{\xi}$,$M_{\Phi_{_{B-L}}}$,$M_{\Phi_{12}}$,$M_{\Phi_3}$, then as the gauge coupling is proportional to the $B-L$ charges, so $Z_{B-L}$ can dominantly decay to the lightest and next to lightest stable particles of the dark sector i.e $Z_{B-L} \rightarrow N_{3R} \overline{N_{3R}}$ or $Z_{B-L} \rightarrow N_{iR} \overline{N_{iR}}$ with i=1,2 ,  (since they have $B-L$ charges $+5$ and $-4$ respectively). Also decay of $Z_{B-L}$ to a pair of SM leptons (having $B-L$ quantum number -1) will be dominant as compared to it's decay to quarks ($B-L$ quantum number $1/3$). But alternatively if $M_{Z_{B-L}} >$ $M_{\xi}$,$M_{\Phi_{_{B-L}}}$,$M_{\Phi_{12}}$,$M_{\Phi_3}$ then the total decay width of $Z_{B-L}$ significantly increases as it additionally decays to $\phi_3 \phi^{*}_3$ , $\phi_{12} \phi^{*}_{12}$, $\xi^{\pm \pm} \xi^{\mp \mp}$, $\xi^{\pm} \xi^{\mp}$, $\xi^{0} \xi^{0*}$ and $\phi_{_{B-L}} \phi^{*}_{_{B-L}}$. 
		
		As already mentioned, this model not only contains the TeV scale $B-L$ gauge boson but also has a triplet scalar $\xi$ with $M_{\xi} \lesssim 1$TeV. So depending on the relative magnitudes of $M_{Z_{B-L}}$ and $M_{\xi^{\pm \pm}}$, the production cross-section of $\xi^{\pm \pm}$ and $Z_{B-L}$ will vary. The collider search for this triplet scalar in various models have been studied in many papers~\cite{Chun:2019hce,Padhan:2019jlc,Akeroyd:2005gt,Huitu:1996su,Hektor:2007uu,Mitra:2016wpr}.
		\begin{figure}[h!]
			\centering
			\includegraphics[width = 75mm]{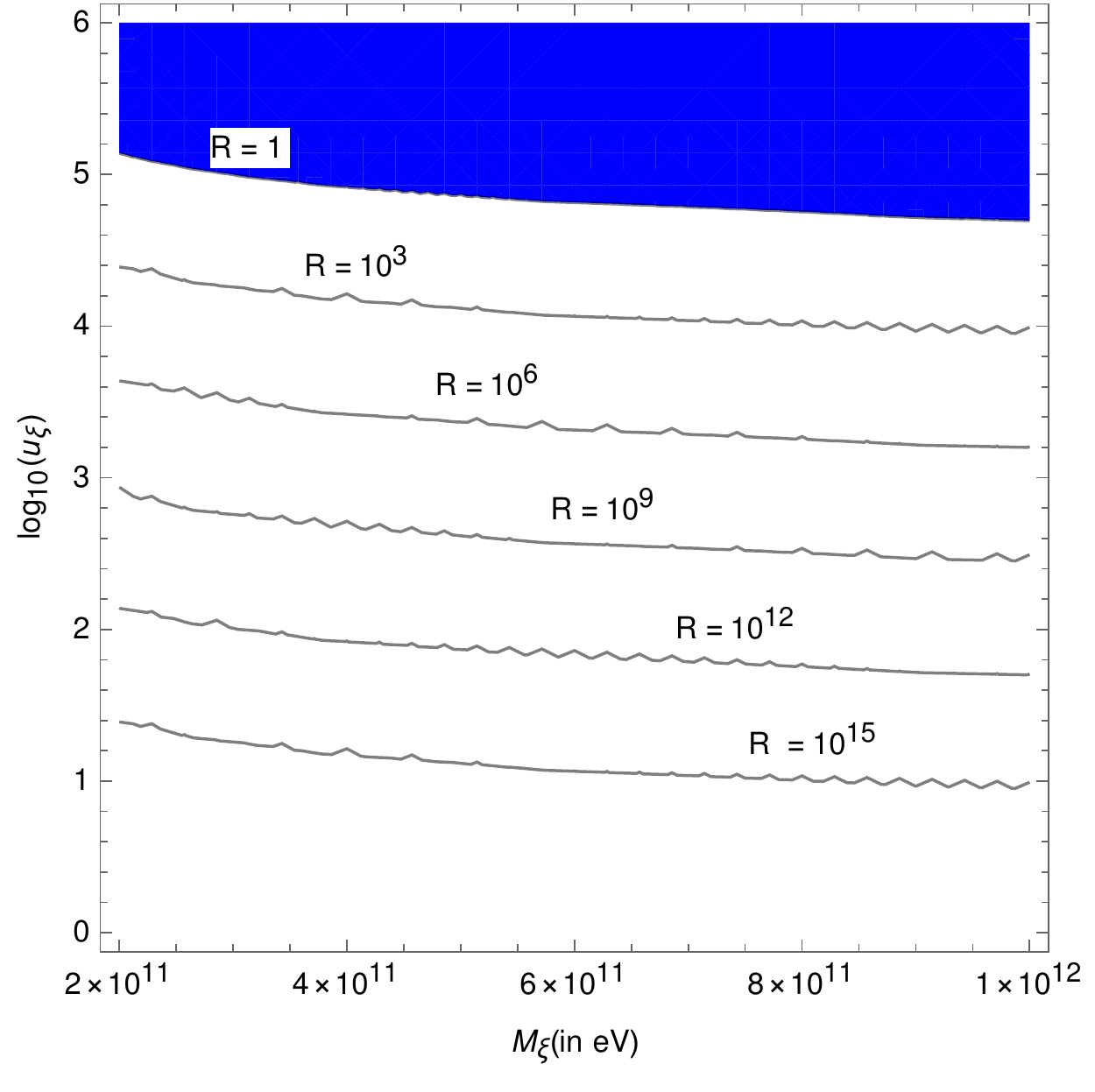}
			\caption{\footnotesize{ The contour for R in the plane of $log_{10}(u_{\xi})$ and $M_\xi$. }}
			\label{Xi_decay_contour}
		\end{figure} 
		
		If $M_{Z_{B-L}} > 2M_{\xi^{\pm \pm}}$, then at LHC ${\xi^{\pm \pm}}$ particles can be pair produced via $Z_{B-L}$ decay with a significant branching fraction. But if $M_{Z_{B-L}} < 2M_{\xi^{\pm \pm}}$, then at LHC ${\xi^{\pm \pm}}$ particles can be produced via Drell-Yan process ($q \overline{q} \rightarrow \xi^{\pm \pm} \xi^{\mp \mp} $). Once $\xi^{\pm \pm}$ are produced ,then it can decay to different SM particles which can be studied at present and future colliders. The $\xi^{\pm \pm}$ particle can decay to two like-sign charged leptons ($l^{+}_\alpha l^{+}_\beta , \alpha ,\beta = e,\mu,\tau$) or to $W^{+} W^{+}$. Though it can also decay to $W^{+} \xi^{+}$, but this decay rate [ $\Gamma{(\xi^{++} \rightarrow W^{+}\xi^{+})}$ ] is phase space suppressed since mass of $\xi^{+}$ is of the same order of mass of $\xi^{++}$. It is worth mentioning that for small $u_{\xi}$, $\Gamma(\xi^{++} \rightarrow l^{+}_{\alpha}l^{+}_{\beta})$ dominates over $\Gamma{(\xi^{++} \rightarrow W^{+}W^{+})}$ as $\Gamma(\xi^{++} \rightarrow l^{+}_{\alpha}l^{+}_{\beta})$ varies inversely with $u^2_{\xi}$ but $\Gamma{(\xi^{++} \rightarrow W^{+}W^{+})}$ varies directly with $u^2_{\xi}$. This is evident from the partial decay widths which are given as:
		\begin{equation}
		\Gamma(\xi^{++} \rightarrow l^{+}_{\alpha}l^{+}_{\beta})= \frac{M_{\xi^{++}}}{4 \pi u^2_{\xi}(1+\delta_{\alpha \beta})} {\vert (M_{\nu})_{\alpha \beta}} \vert^2
		\end{equation}
		and 
		\begin{equation}
		\Gamma(\xi^{++}\rightarrow W^{+}W^{+})=g^4 u^2_{\xi} M^3_{\xi^{++}} \Big[1-4\Big(\frac{M_W}{M_{\xi^{++}}}\Big)^2\Big]^{\frac{1}{2}}\Big[1-4\Big(\frac{M_W}{M_{\xi^{++}}}\Big)^2+\Big(\frac{M_W}{M_{\xi^{++}}}\Big)^4\Big]
		\end{equation}
		This can be well analysed by plotting contours of the ratio
		\begin{equation}
		R = \frac{\Gamma(\xi^{++} \rightarrow l^{+}_{\alpha}l^{+}_{\beta})}{\Gamma(\xi^{++}\rightarrow W^{+}W^{+})}
		\end{equation}
		
		in the plane of $M_{\xi^{++}}$ vs $u_{\xi}$ as shown in the Fig~\ref{Xi_decay_contour}. Only in the region above $R=1$ contour where $R < 1$ $\xi$ dominantly decays to $W^+ W^+$ while for all other regions the dilepton decay is the dominant one. 
		This like-sign dilepton channel of $\xi^{++}$ is almost background free and can be seen at LHC. The mass of $\xi^{++}$ is approximately about the invariant mass of the two like sign leptons. The small SM background for this dilepton signature coming from two Z-boson decay can be removed by proper selection of cuts.

		The singly charged scalar particles $\xi^{\pm}$ can be produced along with the doubly charged scalars through the Drell-Yan process ($q \overline{q'} \rightarrow \xi^{\mp \mp} \xi^{\pm} $) mediated via the charged weak gauge boson $W^{\pm*}$. Also there can be pair production of $\xi^{\pm}$ similar to $\xi^{\pm \pm}$ through Drell-Yan process ($q \overline{q} \rightarrow \xi^{\mp} \xi^{\pm}$) mediated by $\gamma^{*}$,$Z^{*}$,$Z^{*}_{B-L}$. Once produced, the decay of $\xi^{\pm}$ can dominantly ocuur through the channel $\xi^{+} \rightarrow l^+ + \nu$. But since the neutrinos are invisible at the detector, the decay of $\xi^{\pm}$, produced through the channel $q \overline{q'} \rightarrow \xi^{\mp \mp} \xi^{\pm} $ mediated via the the charged weak gauge boson $W^{\pm*}$, will lead to a three lepton final state ($l^{\pm}l^{\pm}l^{\mp}$). On the other hand, the decay of $\xi^{\pm}$ , produced through the channel $q \overline{q} \rightarrow \xi^{\mp} \xi^{\pm}$ will lead to a two lepton final state ($l^{\pm}l^{\mp}$). However in both the cases we will have large SM background, so with proper application of cuts, these events can be studied at LHC.          
		
		\section{Conclusion}
		In this paper we studied a gauged $U(1)_{B-L}$ extension of a TeV scale type-II seesaw. We implemented it by introducing two scalar triplets $\Delta$ ($M_\Delta \simeq 10^{14}$ GeV) and $\xi$  ($M_\xi \leq$ 1 TeV ) with $M_\xi << M_\Delta$. Even though there is orders of magnitude difference between the masses of these two scalars but their contribution to the neutrino mass is identical because of the small mixing between them which arises at TeV scale when $U(1)_{B-L}$ is broken by the vev of the singlet scalar $\Phi_{B-L}$. $\Delta$ being super heavy is decoupled from the low energy effective theory however the decay of leptophillic $\xi^{++}$ is almost background free and can be studied at LHC. To make the theory free from anomalies we introduced three right handed neutrino fields $\nu_{R_{i}} (i=1, 2, 3)$ which have charges under $U(1)_{B-L}$ as -4, -4, +5 respectively. The $U(1)_{B-L}$ charges of right handed neutrinos precludes any coupling with the SM fermions. The lightest one among these right handed neutrinos is the DM candidate of the model whose stability is guaranteed by a remnant $Z_2$ symmetry after $U(1)_{B-L}$ breaking. We showed a combined parameter space which allows both the relic and direct detection bounds. The model under consideration here explains the smallness of neutrino masses as well as DM. To satisfy the relic density constraint of the DM we have considered only the contribution coming from annihilation cross-sections of the DM. We will study the contribution of co-annihilation channels in a future project.
		
		

\begin{thebibliography}{99}
			
			\bibitem{Bertone:2004pz} 
			G.~Bertone, D.~Hooper and J.~Silk,
			Phys.\ Rept.\  {\bf 405}, 279 (2005)
			[hep-ph/0404175].
			
			\bibitem{Feng:2010gw} 
			J.~L.~Feng,
			Ann.\ Rev.\ Astron.\ Astrophys.\  {\bf 48}, 495 (2010)
			[arXiv:1003.0904 [astro-ph.CO]].
			
			\bibitem{Hinshaw:2012aka} 
			G.~Hinshaw {\it et al.} [WMAP Collaboration],
			Astrophys.\ J.\ Suppl.\  {\bf 208}, 19 (2013)
			[arXiv:1212.5226 [astro-ph.CO]].
			
			\bibitem{Aghanim:2018eyx} 
			N.~Aghanim {\it et al.} [Planck Collaboration],
			arXiv:1807.06209 [astro-ph.CO].
			
			\bibitem{solar-expt} Q.R.~Ahmed {\it et al} (SNO Collaboration), 
			Phys.\ Rev.\ Lett. {\bf 89}, 011301-011302 (2002);
			J.N.~Bahcall and C.~Pena-Garay, [arXiv:hep-ph/0404061].
			
			\bibitem{atmos-expt} S.~Fukuda {\it et al} (Super-Kamiokande
			Collaboration), Phys.\ Rev.\ Lett. {\bf 86}, 5656 (2001).
			
			\bibitem{kamland} K.~Eguchi {\it et al} (KamLAND collaboration),
			Phys.~Rev.~Lett. {\bf 90}, 021802 (2003).
			
			\bibitem{Weinberg:1979sa} 
			S.~Weinberg,
			Phys.\ Rev.\ Lett.\  {\bf 43}, 1566 (1979).
			doi:10.1103/PhysRevLett.43.1566
			
			\bibitem{Ma:1998dn} 
			E.~Ma,
			Phys.\ Rev.\ Lett.\  {\bf 81}, 1171 (1998)
			doi:10.1103/PhysRevLett.81.1171
			[hep-ph/9805219].
			
			\bibitem{Minkowski:1977sc} 
			P.~Minkowski,
			Phys.\ Lett.\  {\bf 67B}, 421 (1977).
			doi:10.1016/0370-2693(77)90435-X
			
			\bibitem{GellMann:1980vs} 
			M.~Gell-Mann, P.~Ramond and R.~Slansky,
			Conf.\ Proc.\ C {\bf 790927}, 315 (1979)
			[arXiv:1306.4669 [hep-th]].
			
			\bibitem{Mohapatra:1979ia} 
			R.~N.~Mohapatra and G.~Senjanovic,
			Phys.\ Rev.\ Lett.\  {\bf 44}, 912 (1980).
			
			\bibitem{Schechter:1980gr} 
			J.~Schechter and J.~W.~F.~Valle,
			Phys.\ Rev.\ D {\bf 22}, 2227 (1980).
			
			\bibitem{Mohapatra:1980yp} 
			R.~N.~Mohapatra and G.~Senjanovic,
			Phys.\ Rev.\ D {\bf 23}, 165 (1981).
			
			\bibitem{Lazarides:1980nt} 
			G.~Lazarides, Q.~Shafi and C.~Wetterich,
			Nucl.\ Phys.\ B {\bf 181}, 287 (1981).
			
			\bibitem{Wetterich:1981bx} 
			C.~Wetterich,
			Nucl.\ Phys.\ B {\bf 187}, 343 (1981).
			
			\bibitem{Schechter:1981cv} 
			J.~Schechter and J.~W.~F.~Valle,
			Phys.\ Rev.\ D {\bf 25}, 774 (1982).
			
			\bibitem{Brahmachari:1997cq} 
			B.~Brahmachari and R.~N.~Mohapatra,
			Phys.\ Rev.\ D {\bf 58}, 015001 (1998)
			[hep-ph/9710371].
			
			\bibitem{Foot:1988aq} 
			R.~Foot, H.~Lew, X.~G.~He and G.~C.~Joshi,
			Z.\ Phys.\ C {\bf 44}, 441 (1989).
			
			\bibitem{McDonald:2007ka} 
			J.~McDonald, N.~Sahu and U.~Sarkar,
			JCAP {\bf 0804}, 037 (2008)
			[arXiv:0711.4820 [hep-ph]].
			
			\bibitem{Majee:2010ar} 
			S.~K.~Majee and N.~Sahu,
			Phys.\ Rev.\ D {\bf 82}, 053007 (2010)
			[arXiv:1004.0841 [hep-ph]].
			
			\bibitem{Gu:2009hu} 
			P.~H.~Gu, H.~J.~He, U.~Sarkar and X.~m.~Zhang,
			Phys.\ Rev.\ D {\bf 80}, 053004 (2009)
			[arXiv:0906.0442 [hep-ph]].
			
			\bibitem{Perez:2008ha} 
			P.~Fileviez Perez, T.~Han, G.~y.~Huang, T.~Li and K.~Wang,
			Phys.\ Rev.\ D {\bf 78}, 015018 (2008)
			[arXiv:0805.3536 [hep-ph]].
			
			\bibitem{Chun:2003ej} 
			E.~J.~Chun, K.~Y.~Lee and S.~C.~Park,
			Phys.\ Lett.\ B {\bf 566}, 142 (2003)
			[hep-ph/0304069].
			\bibitem{Chun:2019hce} 
			E.~J.~Chun, S.~Khan, S.~Mandal, M.~Mitra and S.~Shil,
			arXiv:1911.00971 [hep-ph].
			
			\bibitem{Padhan:2019jlc} 
			R.~Padhan, D.~Das, M.~Mitra and A.~Kumar Nayak,
			arXiv:1909.10495 [hep-ph].
			
			
			\bibitem{Akeroyd:2005gt} 
			A.~G.~Akeroyd and M.~Aoki,
			Phys.\ Rev.\ D {\bf 72}, 035011 (2005)
			[hep-ph/0506176].
			
			\bibitem{Huitu:1996su} 
			K.~Huitu, J.~Maalampi, A.~Pietila and M.~Raidal,
			Nucl.\ Phys.\ B {\bf 487}, 27 (1997)
			[hep-ph/9606311].
			
			\bibitem{Hektor:2007uu} 
			A.~Hektor, M.~Kadastik, M.~Muntel, M.~Raidal and L.~Rebane,
			Nucl.\ Phys.\ B {\bf 787}, 198 (2007)
			[arXiv:0705.1495 [hep-ph]].
			
			\bibitem{Mitra:2016wpr}
			M.~Mitra, S.~Niyogi and M.~Spannowsky,
			Phys.\ Rev.\ D {\bf 95} (2017) no.3,  035042
			[arXiv:1611.09594 [hep-ph]].
			
			
			
			
			
			
			
			\bibitem{Montero:2007cd} 
			J.~C.~Montero and V.~Pleitez,
			Phys.\ Lett.\ B {\bf 675}, 64 (2009)
			[arXiv:0706.0473 [hep-ph]].
			
			\bibitem{Sanchez-Vega:2014rka} 
			B.~L.~Sánchez-Vega, J.~C.~Montero and E.~R.~Schmitz,
			Phys.\ Rev.\ D {\bf 90}, no. 5, 055022 (2014)
			[arXiv:1404.5973 [hep-ph]].
			
			\bibitem{Ma:2014qra} 
			E.~Ma and R.~Srivastava,
			Phys.\ Lett.\ B {\bf 741}, 217 (2015)
			[arXiv:1411.5042 [hep-ph]].
			
			\bibitem{Sanchez-Vega:2015qva} 
			B.~L.~Sánchez-Vega and E.~R.~Schmitz,
			Phys.\ Rev.\ D {\bf 92}, 053007 (2015)
			[arXiv:1505.03595 [hep-ph]].
			
			\bibitem{Ma:2015mjd} 
			E.~Ma, N.~Pollard, R.~Srivastava and M.~Zakeri,
			Phys.\ Lett.\ B {\bf 750}, 135 (2015)
			[arXiv:1507.03943 [hep-ph]].
			
			\bibitem{Gu:2019ird} 
			P.~H.~Gu,
			arXiv:1907.11557 [hep-ph].
			
			
			
			\bibitem{Patra:2016ofq} 
			S.~Patra, W.~Rodejohann and C.~E.~Yaguna,
			JHEP {\bf 1609}, 076 (2016)
			[arXiv:1607.04029 [hep-ph]].
			
			\bibitem{Nanda:2017bmi} 
			D.~Nanda and D.~Borah,
			Phys.\ Rev.\ D {\bf 96}, no. 11, 115014 (2017)
			[arXiv:1709.08417 [hep-ph]].
			
			\bibitem{pbpalbook} See for a text book reference: 
			P. Pal, {\it An Introductory Course of Particle Physics. CRC Press, Taylor \& Francis Group, 2014 (1st Edition)}.
			
			\bibitem{Belanger:2008sj} 
			G.~Belanger, F.~Boudjema, A.~Pukhov and A.~Semenov,
			Comput.\ Phys.\ Commun.\  {\bf 180}, 747 (2009)
			[arXiv:0803.2360 [hep-ph]].
			
			
			
			\bibitem{Aaboud:2017buh} 
			M.~Aaboud {\it et al.} [ATLAS Collaboration],
			JHEP {\bf 1710}, 182 (2017)
			[arXiv:1707.02424 [hep-ex]].
			
			\bibitem{Ellis:2000ds} 
			J.~R.~Ellis, A.~Ferstl and K.~A.~Olive,
			Phys.\ Lett.\ B {\bf 481}, 304 (2000)
			[hep-ph/0001005].
			
			\bibitem{Aprile:2018dbl} 
			E.~Aprile {\it et al.} [XENON Collaboration],
			Phys.\ Rev.\ Lett.\  {\bf 121}, no. 11, 111302 (2018)
			[arXiv:1805.12562 [astro-ph.CO]].
			
			\bibitem{Emam:2007dy} 
			W.~Emam and S.~Khalil,
			Eur.\ Phys.\ J.\ C {\bf 52}, 625 (2007)
			[arXiv:0704.1395 [hep-ph]].
			
			\bibitem{Melfo:2011nx} 
			A.~Melfo, M.~Nemevsek, F.~Nesti, G.~Senjanovic and Y.~Zhang,
			Phys.\ Rev.\ D {\bf 85}, 055018 (2012)
			[arXiv:1108.4416 [hep-ph]].
			
			
			
			
		\end{thebibliography}
	\end{document}